\documentclass{aa}  

\usepackage[varg]{txfonts}
\usepackage{graphicx}
\usepackage[]{natbib}
\bibpunct{(}{)}{;}{a}{}{,} 
\usepackage[usenames,dvipsnames,svgnames,table]{xcolor}
\usepackage{array,longtable}
\usepackage{verbatim}
\usepackage{moreverb}
\usepackage{pdflscape}
\usepackage{afterpage}
\usepackage{enumitem}
\usepackage{datetime}
\usepackage{ragged2e}
   \renewcommand{\fs}{1}
\mathchardef\mhy="2D

\definecolor{gold}{HTML}{FFD700}
\definecolor{pink}{HTML}{DC143C}
\definecolor{violet}{HTML}{0000FF}
\definecolor{WRblue}{HTML}{1E90FF}

\begin{document}

\title{Supergiants and their shells in young globular clusters}
\titlerunning{Supergiants and shells in young GCs}
\authorrunning{Dorottya Sz\'ecsi et al.}

   \author{Dorottya Sz\'ecsi$^{1,2}$\thanks{\textsl{Email: dorottya.szecsi@gmail.com}}
\and Jonathan Mackey$^{3,4,5}$
\and Norbert Langer$^5$
}

   \institute{$^1$~Astronomical Institute of the Czech Academy of Sciences, Fri\v{c}ova 298, 25165 Ond\v{r}ejov, Czech Republic\\
   $^2$~Institute of Gravitational Wave Astronomy and School of Physics and Astronomy, University of Birmingham, Edgbaston, Birmingham B15 2TT, UK\\
   $^3$~Dublin Institute for Advanced Studies, School of Cosmic Physics, 31 
   Fitzwilliam Place, Dublin 2, Ireland\\
   $^4$~I.\ Physikalisches Institut, Universit\"at zu K\"oln, Z\"ulpicher 
   Stra\ss{}e 77, 50937 K\"oln, Germany\\
   $^5$~Argelander-Institut f\"ur Astronomie der Universit\"at Bonn, Auf dem H\"ugel 71, 53121 Bonn, Germany}

   \date{Accepted on Friday 10th November, 2017}

\abstract{Anomalous surface abundances are observed in a fraction of the low-mass stars of Galactic globular clusters, that may originate from hot-hydrogen-burning products ejected by a previous generation of massive stars. } 
{We present and investigate a scenario in which the second generation of polluted low-mass stars can form in shells around cool supergiant stars within a young globular cluster.} 
{Simulations of low-metallicity massive stars (M$_{\rm i}$~$\sim$~150$-$600~M$_{\odot}$) show that both core-hydrogen-burning cool supergiants and hot ionizing stellar sources are expected to be present simulaneously in young globular clusters. Under these conditions, photoionization-confined shells form around the supergiants. We simulate such a shell, investigate its stability and analyse its composition.} 
{We find that the shell is gravitationally unstable on a timescale that is shorter than the lifetime of the supergiant, and the Bonnor-Ebert mass of the overdense regions is low enough to allow star formation. Since the low-mass stellar generation formed in this shell is made up of the material lost from the supergiant, its composition necessarily reflects the composition of the supergiant wind. We show that the wind contains hot-hydrogen-burning products, and that the shell-stars therefore have very similar abundance anomalies that are observed in the second generation stars of globular clusters. Considering the mass-budget required for the second generation star-formation, we offer two solutions. Either a top-heavy initial mass function is needed with an index of $-$1.71..$-$2.07. Alternatively, we suggest the shell-stars to have a truncated mass distribution, and solve the mass budget problem by justifiably accounting for only a fraction of the first generation.
} 
{Star-forming shells around cool supergiants could form the second generation of low-mass stars in Galactic globular clusters. Even without forming a photoionizaton-confined shell, the cool supergiant stars predicted at low-metallicity could contribute to the pollution of the interstellar medium of the cluster from which the second generation was born. Thus, the cool supergiant stars should be regarded as important contributors to the evolution of globular clusters. 
} 

%

   \maketitle
%


\section{Introduction}\label{sec:Introduction-shell}

Globular clusters (GC) are found in the halo of the Milky Way orbiting around the Galactic core. They are generally composed of old, low-mass stars bound together by gravity. The composition of these stars may vary between clusters, but in average, GCs have subsolar metallicity \citep[Z,][]{Gratton:2004,Harris:2010}. GCs are under intensive investigation for many reasons. 
Their stars are so old that they constrain the minimum age of the universe.
Additionally, their stars are both coeval and equidistant, thereby providing natural laboratories for stellar evolution.

One of the most intriguing open questions concerning GCs is the so-called abundance anomalies \citep{Yong:2003,DaCosta:2013}. Light element abundances such as O and Na anticorrelate with each other: if O is depleted in a star, then Na is enhanced. The same is observed for the proton-capture isotopes of Mg and Al: if Mg is depleted in a star, then Al is enhanced.
Moreover, with the Al-abundance increasing, the ratio of the $^{24}$Mg isotope to the total Mg is decreasing, the $^{25}$Mg is slightly decreasing and the $^{26}$Mg is considerably increasing in the observed GC stars. This is consistent with the interpretation that one generation of stars has been polluted by nuclear burning products produced at very high temperatures \citep[>6$\cdot$10$^7$~K, ][]{Ventura:2011}.
The nucleosynthetic processes that can increase Na and Al while destroying O and Mg (as well as creating the Mg-isotopic ratios observed) are the Ne-Na chain and the Mg-Al chain \citep{Burbidge:1957}, respectively. These burning chains are side-reactions of the CNO-cycle, the main hydrogen-burning process in \textsl{massive} stars. Consequently, there must have been at least one population of massive (and/or intermediate-mass) stars born in the early epochs of the GC's life. These massive stars are already dead, but their nuclear imprint is what we observe today as anomalous abundance patterns in the second generation of low-mass stars. The question is then: how did the pollution happen, i.e.\ how did massive stars lose the amount of nuclear-processed material, and how did this material end up in some of the low-mass stars?

According to the most commonly accepted explanation, the interstellar medium (ISM) had been polluted by hydrogen-burning products from massive stars, and the second generation of stars were born from the polluted ISM \citep{Decressin:2007,DErcole:2008}. Alternatively, low-mass stars could accrete the ISM during a long pre-main sequence phase \citep{Bastian:2013}. In both cases, an astrophysical source -- a polluter -- is needed. This source, a population of massive or intermediate-mass stars, should only produce hydrogen-burning products (including helium), since no traces of helium burning products or supernova ejecta are observed. Additionally, the polluter should eject the material slowly enough for it to stay inside the gravitational potential well of the GC. This condition excludes fast winds of massive OB stars or Wolf-Rayet stars unless the fast winds are shocked and can cool efficiently before leaving the cluster \citep[cf.][]{Wunsch:2016}.

Several astrophysical scenarios were proposed that fulfill the conditions above. Asymptotic giant branch stars could eject their hot bottom burning products \citep{Ventura:2001,DErcole:2008}. Fast rotating massive stars that are close to the breakup rotation could eject core burning products \citep{Decressin:2007,Tailo:2015}. Supermassive (10\,000~M$_{\odot}$) stars could pollute through continuum driven stellar wind \citep{Denissenkov:2014}. In addition, massive binary systems could pollute via non-conservative mass transfer \citep{deMink:2009}. 

\begin{figure}
\centering
\resizebox{1.1\hsize}{!}{\includegraphics[page=1]{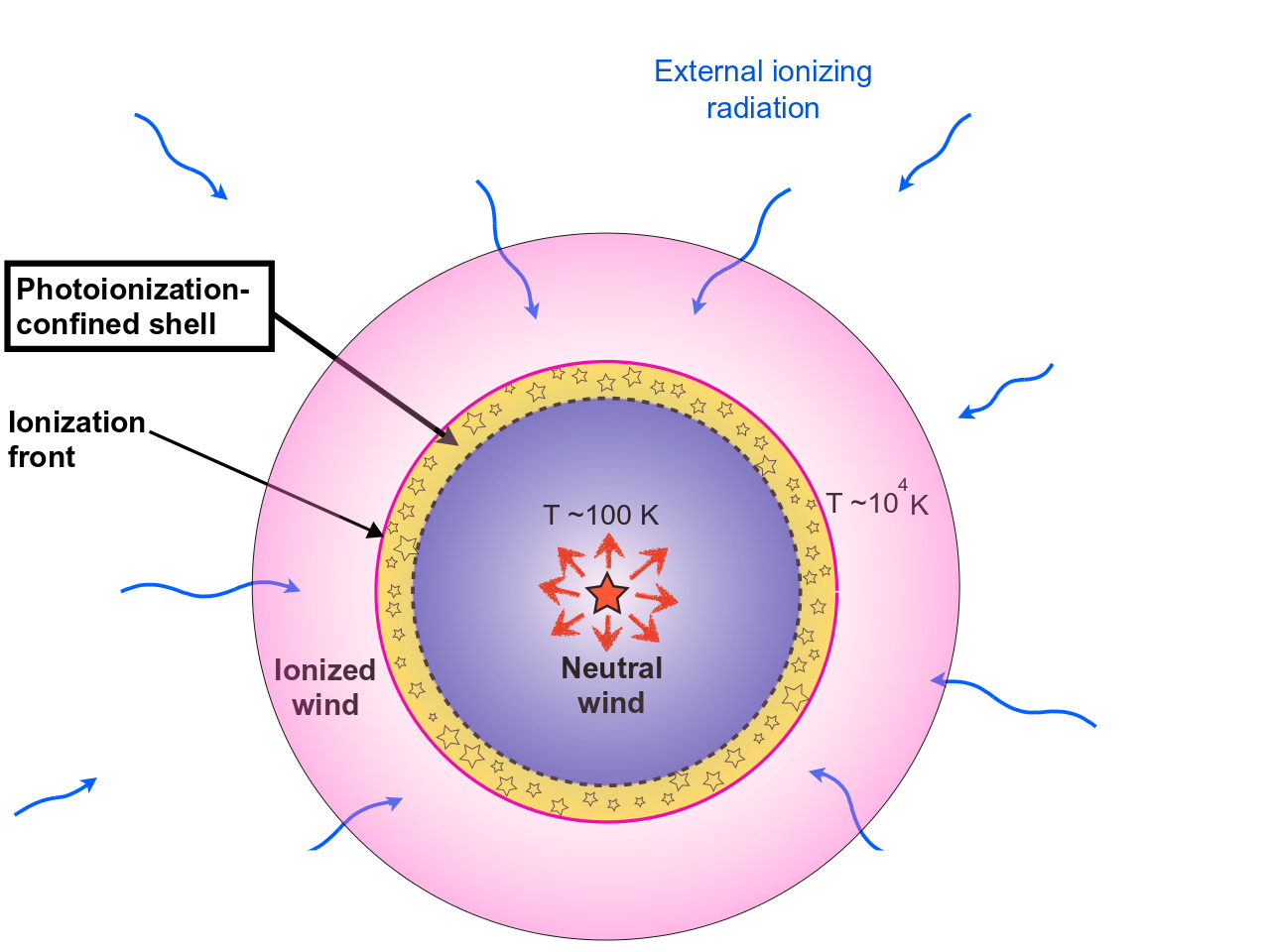}}
\caption{Photoionization-confined shell around a cool supergiant star. The second generation of low-mass stars are formed in the shell. This scenario could be common in the first few million years of the early globular clusters, explaining the pollution of the second generation. This simple drawing serves to present the original idea; as for the nominal values of our model, the shell forms at $r\approx 0.02$~pc from the central star (cf.\ our simulation of a shell in Fig.~\ref{fig:shelldens}). The central supergiant itself has a stellar radius of $\sim$5000~R$_{\odot}$; that is, the supergiant is 170 times smaller in radial dimension than the sphere of the shell. (This figure is derived from fig.~1 of \citet{Mackey:2014}).
}
\label{fig:shell}
\end{figure}

Here we propose a new scenario: low mass stars could be born in photoionization-confined shells around cool supergiant (SG) stars in the young globular clusters, as shown in Fig.~\ref{fig:shell}.
\citet{Szecsi:2015} simulated very massive (80$-$300~M$_{\odot}$) and long-living SGs. These long-living SGs are predicted only to exist at low-Z, because at solar composition the strong mass-loss removes their envelopes and turns them into Wolf--Rayet stars before reaching the SG branch. Moreover, the very massive, metal-poor SGs form \textsl{before} the hydrogen is exhausted in the core \citep[this is due to envelope inflation, cf.][]{Sanyal:2015}. Core-hydrogen-burning cool supergiants spend 0.1-0.3 Myr in the SG branch. During this time, they lose a large amount of mass (up to several hundred M$_{\odot}$ in the case of a 600~M$_{\odot}$ star, as we show below). This mass lost in the SG wind has undergone nuclear burning and shows similar abundance variations to those observed in GC stars. 

Photoionization-confined shells can be present around cool supergiants at the interface of ionized and neutral material, as shown by \citet{Mackey:2014}. The shell can contain as much as 35\% of the mass lost in the stellar wind. 
The main condition for forming a photoionization-confined shell is that the SG has a cool and slow wind and is surrounded by strong sources of Lyman-continuum radiation.
These conditions may have been fulfilled at the time when Galactic globular clusters were born. Evolutionary simulations of low-Z massive stars by \citet{Szecsi:2015} predict that both supergiant stars and compact hot stars develop at the same time. The latter are fast rotating, hot and luminous massive stars that
emit a huge number of Lyman-continuum photons. The slowly rotating stars, on the other hand, evolve to be cool red or yellow SGs. Thus, the condition required by \citet{Mackey:2014} about SGs and ionizing sources close to each other may have been common in the first few million years of a GC's life. 
Consequently, photoionization-confined shells could form there, too.

This work is organized as follows. In Sect.~\ref{sec:SGinyoungGC} we present the evolution of the models that become core-hydrogen-burning cool SG stars, and discuss the composition of their winds. In Sect.~\ref{sec:SGshells} we introduce the star-forming supergiant shell scenario, and show that in the environment of the young globular clusters, it is possible to form low-mass stars in a supergiant shell from the material ejected by the SG's wind. In Sect.~\ref{sec:discuu} we discuss the mass budget of our scenario, as well as the amount of helium predicted in the second generation. In Sect.~\ref{sec:conclusionshell} we summarize the work. 

\section{Supergiants in young GCs}\label{sec:SGinyoungGC}

\subsection{The evolution of core-hydrogen-burning cool SGs}\label{sec:evolution}

The first generation of stars in the young GCs almost certainly contained massive stars. We see massive stars forming in young massive clusters (YMC) today \citep{Longmore:2014}. YMCs are theoretically similar to the young GCs and are thought to become GC-like objects eventually 
\citep[e.g.][]{Brodie:2006,Mucciarelli:2014,Andersen:2016}.

\begin{figure}
\centering
\resizebox{0.5\hsize}{!}{\includegraphics[width=0.2\columnwidth,angle=0]{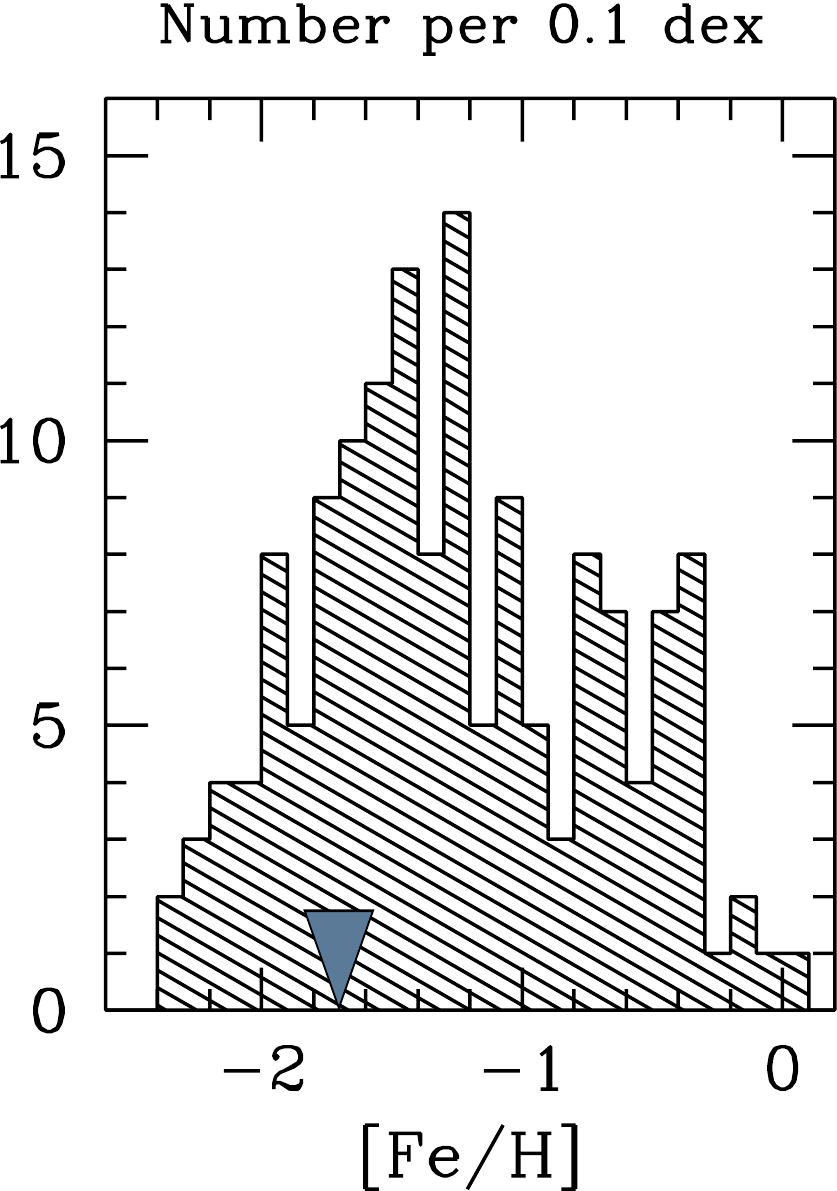}}
\caption{Number of GCs at a given metallicity. The figure is taken from \citet[][]{Harris:2010}, and shows the distribution of 157 GCs with measured [Fe/H] value. We apply a metallicity of [Fe/H]=$-$1.7 (marked in the figure) to model the first generation of massive stars in GCs.
}
\label{fig:histo}
\end{figure}

The massive stars of this first generation must have had the same metallicity that we observe today in the low-mass GC stars.  The metallicity distribution of GCs in the Galaxy is shown in Fig.~\ref{fig:histo}. It is a broad and bi-modal distribution with a large peak at [Fe/H]~$\sim -$1.4 and a smaller peak at $\sim -$0.6 \citep[cf.][]{Gratton:2004,Brodie:2006,Harris:2006,Harris:2010,Forbes:2010}.  While there is recent evidence that a few of the high-metallicity GCs seem to harbor multiple generations too \citep[][]{Schiavon:2017}, here we only consider low-metallicity GCs that are in the large peak, that is between [Fe/H]=$-$1.0 to $-$2.0, because the abundance anomalies seem to be consistently present in almost all of them \citep{Gratton:2004}.

We use the low-metallicity ([Fe/H]=$-$1.7, corresponding to 0.02~Z$_{\odot}$) massive star simulations of \citet{Szecsi:2015} to model the young GC environment and the first generation of massive and very massive stars. However, \citet{Szecsi:2015} do not use an $\alpha$-enhanced mixture \citep[as suggested for GC stars by][see their Table~3]{Decressin:2007}, but a mixture suitable for dwarf galaxies. Therefore, when comparing to observations (in Figs.~\ref{fig:obsNaO}--\ref{fig:obsMgiso}), the initial O, Na, Mg and Al abundance of our models are scaled to the following abundance ratios:
[O/Fe]$_{\rm first}$=0.4,  
[Na/Fe]$_{\rm first}$=$-$0.4, 
[Mg/Fe]$_{\rm first}$=0.6, 
[Al/Fe]$_{\rm first}$=0.2,
approximately matching the observed composition of the first generation of GC stars.

Massive stars at low Z evolve differently from those at Z$_{\odot}$.
Simulations of \citet{Szecsi:2015} predict different evolutionary paths and, consequently, new types of objects present in low-Z environments. One of the predictions at low Z are the core-hydrogen-burning cool supergiant stars. 
These objects start their evolution as O-type stars but, during their main-sequence phase, they expand due to envelope inflation \citep{Sanyal:2015} and become cool SG stars while still burning hydrogen in their cores. 
The cool supergiants in general have a convective envelope because of their low (<10$^4$~K) surface temperature. Envelope convection mixes nuclear products from the burning regions (core or shell) to the surface. Thus, the wind of the cool SG stars contains the products of nuclear burning that is happening in the deeper regions of these stars. In case of core-hydrogen-burning cool supergiants, the nuclear burning products in the wind are, necessarily, hot-hydrogen-burning products. 

Core-hydrogen-burning cool SGs with low metallicity (0.02~Z$_{\odot}$) are predicted at masses higher than M$_{\rm ini}\gtrsim$~80~M$_{\odot}$.
They stay on the SG branch and burn hydrogen for a relatively long time (in some cases, as long as 0.3~Myr, which corresponds to 15\% of their main sequence lifetimes). These objects have a contribution to the chemical evolution of their environments. Such a star could eject several tens, or hundreds, of M$_{\odot}$ through stellar wind mass-loss, the composition of which material being different from that of the circumstellar gas. 

We simulate the cool supergiant phase by applying the mass-loss rate prescription by \citet{Nieuwenhuijzen:1990}, which is a parametrized version of that by \citet{deJager:1988}. The latter has been shown by \citet{Mauron:2011} to be still applicable in the light of new observations of red supergiants. A metallicity-dependence of the wind is implemented as $\dot{M}\sim Z^{0.85}$ according to \citet{Vink:2001}. Thus, the mass-loss recipe we use:
\begin{equation}
\begin{split}
\log\frac{\dot{M}}{M_{\odot}{\rm yr}^{-1}}= 1.42\log (L/L_{\odot}) + 0.16\log (M/M_{\odot}) + \\
+ 0.81\log (R/R_{\odot}) - 15\log(9.6310) + 0.85\log (Z_{\rm ini}/Z_{\odot})\label{eq:nieu}
\end{split}
\end{equation}
This formula is in accordance with the results of \citet{Mauron:2011} who find that the metallicity exponent should be between 0.5 and 1. However, it is important to note that this prescription is based on red SG stars with masses between 8-25~M$_{\odot}$. Since there is no mass-loss rate observed for SG stars with masses of 150-600~M$_{\odot}$, we extrapolate Eq.~\ref{eq:nieu} up to these masses, pointing out that this approach involves large uncertainties. 

\begin{figure}
\centering
\resizebox{\hsize}{!}{\includegraphics[width=0.5\columnwidth,angle=270]{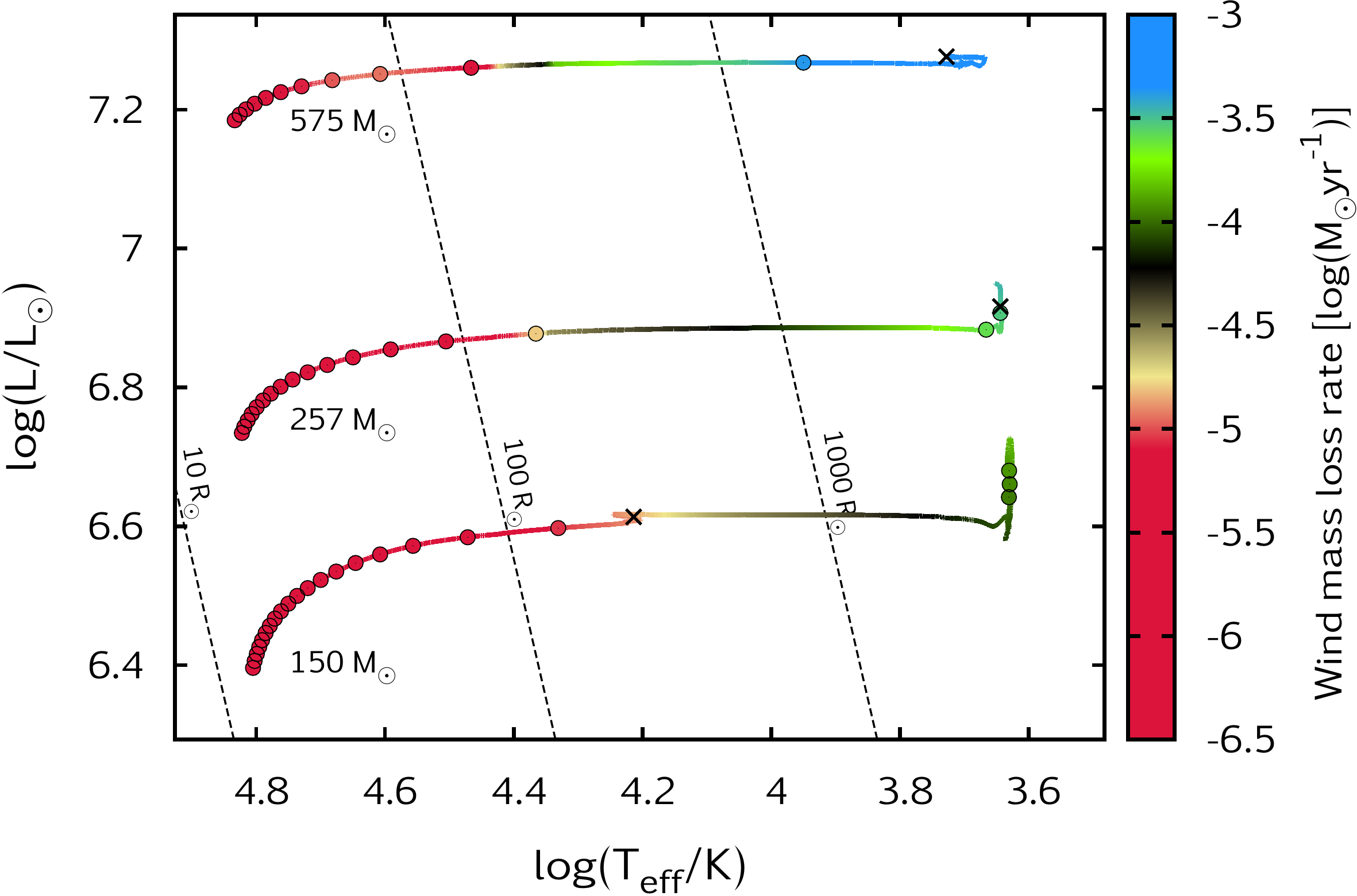}}
\caption{Hertzsprung--Russell diagram of three low-Z evolutionary models that become core-hydrogen-burning SG stars with initial masses of 150, 257 and 575~M$_{\odot}$ and initial rotational velocity of 100~km~s$^{-1}$. Dots in the tracks mark every 10$^5$~years of evolution. Crosses mark the end of the core-hydrogen-burning phase; in case of the model with 575~M$_{\odot}$, the end of the computation.
Theoretical mass-loss rates are colour coded, and dashed lines indicate the radial size of the stars on the diagram. 
}
\label{fig:HR}
\end{figure}

Fig.~\ref{fig:HR} shows the Hertzsprung--Russell diagram of three evolutionary models that become core-hydrogen-burning SG stars
towards the end of their main-sequence evolution.
The models were taken from \citet{Szecsi:2015}, except for the most massive one (M$_{\rm ini}$=575~M$_{\odot}$) which was computed for this work.  
Our simulation of the model with M$_{\rm ini}$=575~M$_{\odot}$ was carried out until the central helium mass-fraction was 0.81, that is, before the end of core hydrogen-burning.
We estimate that until core-hydrogen exhaustion, this model needs about 0.28~Myr of further evolution, thus the total time it spends as a core-hydrogen-burning cool SG is 0.37~Myr. Based on its main-sequence lifetime of 1.56~Myr and the general trend that massive stars spend 90\% of their total life on the main-sequence and 10\% on the post-main-sequence, we expect a post-main-sequence lifetime of $\sim$0.17~Myr. The mass loss in the SG phase can be as high as 10$^{-3}$~M$_{\odot}$~yr$^{-1}$.
It is expected that with this high mass-loss, the model loses its whole envelope during its post-main-sequence lifetime. But even if all its hydrogen-rich layers are lost, it will stay cool. According to \citet[][their fig.~19]{Koehler:2015} the zero-age main-sequence (ZAMS) of pure helium-stars bends toward that of hydrogen-rich stars, crossing it over at $\sim$300~M$_{\odot}$ in the case of models with subsolar (SMC and LMC) composition. Although the exact mass where the crossover of the two ZAMS-lines happens at our sub-SMC metallicity needs to be investigated in the future, the model with M$_{\rm ini}$=575~M$_{\odot}$ (and a total mass of 491~M$_{\odot}$ at the end of our simulation) is most probably above it. Therefore, we do not expect this model to become a hot Wolf--Rayet star after its envelope is lost, but instead to stay cool, and become a helium-rich SG during the remaining evolution. 

The model with M$_{\rm ini}$=257~M$_{\odot}$ from \citet{Szecsi:2015} was followed during its post-main-sequence evolution. Our simulation stops when the central helium mass fraction has decreased to 0.73 during core \textsl{helium}-burning. The model spends 0.26~Myr as a core-hydrogen-burning cool SG (with a radius of $\sim$5000~R$_{\odot}$~$\sim$3.5$\cdot$10$^{14}$~cm), and is expected to spend a total of $\sim$0.25~Myr as a core-helium-burning object. The mass-loss rate is 2.9$\cdot$10$^{-4}$~M$_{\odot}$~yr$^{-1}$ (i.e.~$-$3.5 on a logarithmic scale) in the last computed model. Supposing that this mass-loss rate stays constant until the end of its post-main-sequence lifetime, this model will end up having only 140~M$_{\odot}$. It remains an open question if this model, having lost its hydrogen-rich envelope, would stay cool or would become a hot Wolf--Rayet star. To decide, one would need either to follow the rest of its evolution, or to establish a mass-limit where the helium-ZAMS and the hydrogen-ZAMS cross. 
Since these tasks would require improvements of the code and creating a dense grid of high-mass models, they fall outside of the scope of current work. 
However, given all the uncertainties concerning the mass-loss rates of actual supergiant stars at this mass, it may be that the model never even loses its envelope because the real mass-loss rate is lower than assumed here.


The model with M$_{\rm ini}$=150~M$_{\odot}$ has finished core-helium-burning in our simulation. It spends 0.07~Myr as a core-hydrogen-burning cool SG (during which time its surface does not become cooler than 19~000~K; its largest radius is 182~R$_{\odot}$) and another 0.30~Myr as a core-\textsl{helium}-burning red supergiant (with a surface temperature of $\sim$4250~K and a radius of $\sim$4000~R$_{\odot}$). It has a final mass of 118~M$_{\odot}$, and the mass-loss rate in the last computed model is 8.0$\cdot$10$^{-5}$~M$_{\odot}$~yr$^{-1}$. 
Since core-helium-burning is finished in this model, we know its final surface temperature, as well as its envelope composition: it is a red supergiant at the end of its life, and it has an envelope of about 25~M$_{\odot}$ which is composed of 49.02\% hydrogen, 50.96\% helium and 0.02\% metals. Thus, we know for sure that it stays cool until the end of its life, whereas we could not be sure for the two more massive models discussed above. Moreover, we find no helium-burning side-products at its surface. The reason for this is that the size of the convective core during helium-burning is smaller than that during hydrogen-burning, and the convective envelope of the red supergiant never reaches the layers of the helium-burning. It only mixes the ashes from core-hydrogen-burning and, during the post-main-sequence phase, shell-hydrogen-burning to the surface. As the observed composition of GC stars show no traces of helium-burning products either, we suggest that this SG model, having finished its post-main-sequence evolution while ejecting about 30~M$_{\odot}$ of material polluted with hot-hydrogen-burning products, is a potential source of the pollution in the young GCs.  

\subsection{Composition of the SG wind}

\begin{figure}
\centering
\resizebox{\fs\hsize}{!}{\includegraphics[page=1]{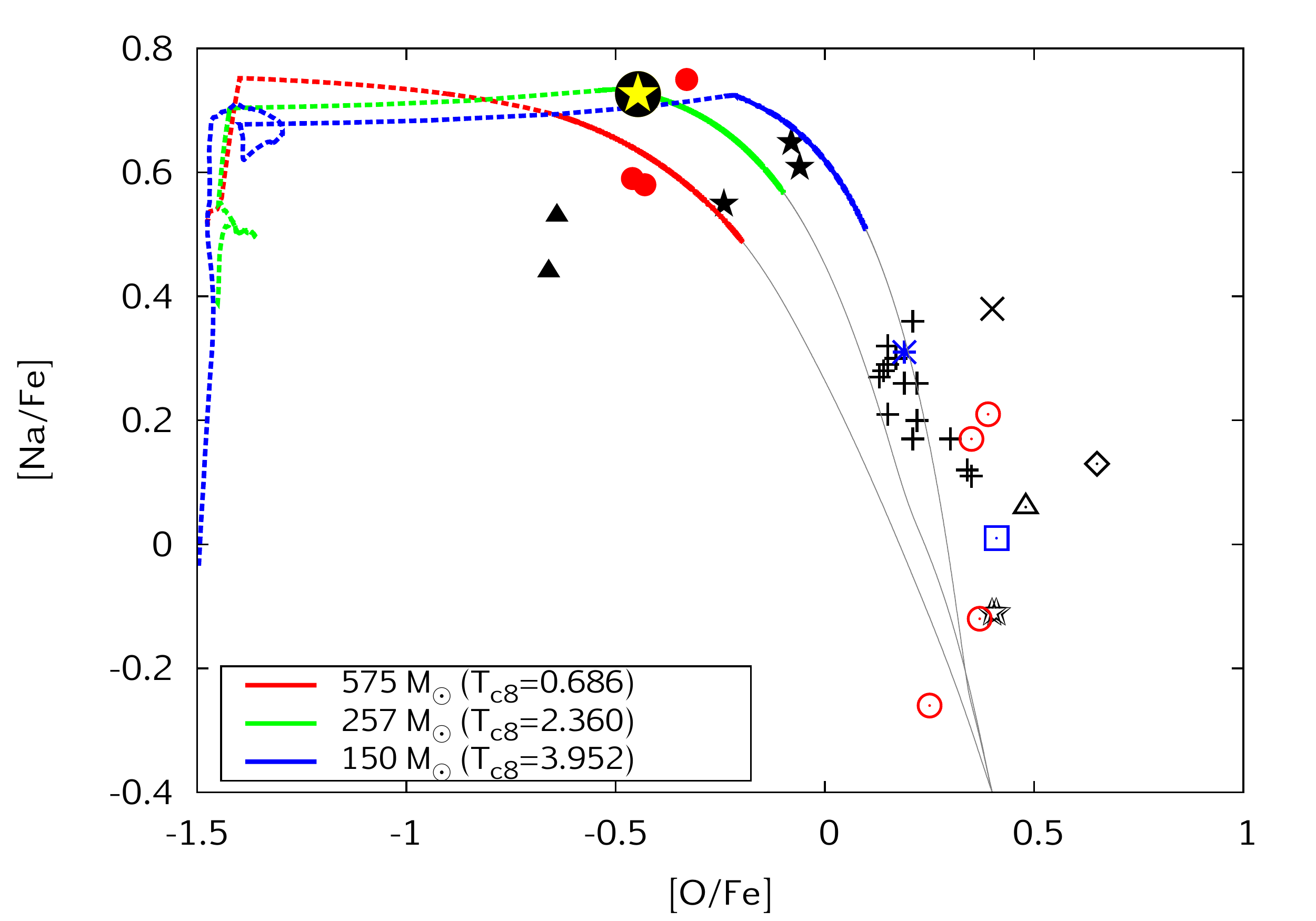}}
\caption{
Theoretical predictions of the wind composition (surface Na abundance as a function of the surface O abundance, in solar Fe units) of three stellar models that become core-hydrogen-burning SGs are plotted with lines. The grey part of the lines correspond to surface compositions at T$_{\rm eff}$>10$^4$~K (i.e.\ the evolution before reaching the SG branch), while the coloured part of the lines show surface composition at T$_{\rm eff}$<10$^4$~K (i.e.\ on the SG branch). When the lines become dashed, they represent the composition of the envelope in the last computed model (i.e.\ deeper layers that could still be lost if the mass-loss rate was higher than assumed here). The evolutionary calculations ended at the core temperatures, T$_{\rm c8}$, given in the legend (units in 10$^8$~K).
The black-yellow star-symbol corresponds to the composition for the simulation presented in Sect.~\ref{sec:comp}.
Observational data of the surface composition of GC stars ($\omega$~Cen red, NGC~6752 black and M~4 blue) are plotted with dots of different colours and shapes, following \citet{Yong:2003}, \citet{DaCosta:2013} and \citet{Denissenkov:2014}. 
Open symbols mark the `primordial' population of stars, that is, those without pollution. Filled symbols mark the `extremely' polluted population of stars. Crosses mark the `intermediate' population stars, that is, those with some but not extreme pollution. For details of the observations and the properties of these categories, we refer to \citet{Yong:2003} and \citet{DaCosta:2013}.
}
\label{fig:obsNaO}
\end{figure}

Core-hydrogen-burning cool SGs have a convective envelope that mixes the hydrogen-burning products from the interior to the surface. The strong stellar wind then removes the surface layers. To calculate the composition of the ejecta, we need to sum over the surface composition of the evolutionary models. Fig.~\ref{fig:obsNaO} shows the surface Na abundance as a function of the surface O abundance of the three models presented above (in Fig.~\ref{fig:HR}). During their SG phase, the surface composition of our models cover the area where the most extremely polluted population of GC stars are found. This means that if low-mass stars form from the material lost by the SG directly (i.e.\ without mixing the ejecta with pristine gas), this second generation of low-mass stars would be observed as part of the extremely polluted population (cf.\ Sect.~\ref{sec:comp}). In case, however, if the material lost via the slow SG wind is mixed with non-polluted gas, the second generation of low-mass stars could possibly reflect the composition of the so-called intermediate population \citep[i.e.\ those stars that show some traces of pollution, compared to a not-polluted, primordial population, as explained by][]{DaCosta:2013}.

Since the mass-loss rates of our models are uncertain, it is worth investigating how a higher mass-loss rate would influence the ejecta composition. Therefore, we also plotted the composition of the envelope in the last model in Fig.~\ref{fig:obsNaO}. With a higher mass-loss rate (or, in the case of the two most massive models, during the remaining evolutionary time), deeper layers could be lost in the wind, contributing to the extremely polluted generation with very low [O/Fe] (<$-$1) and very high [Na/Fe] ($\sim$0.7).  
Deep inside the envelope, the Na abundance drops suddenly because the high temperature ($\gtrsim$0.8$\cdot$10$^{8}$~K) destroys the Na. 

\begin{figure}
\centering
\resizebox{\fs\hsize}{!}{\includegraphics[page=2]{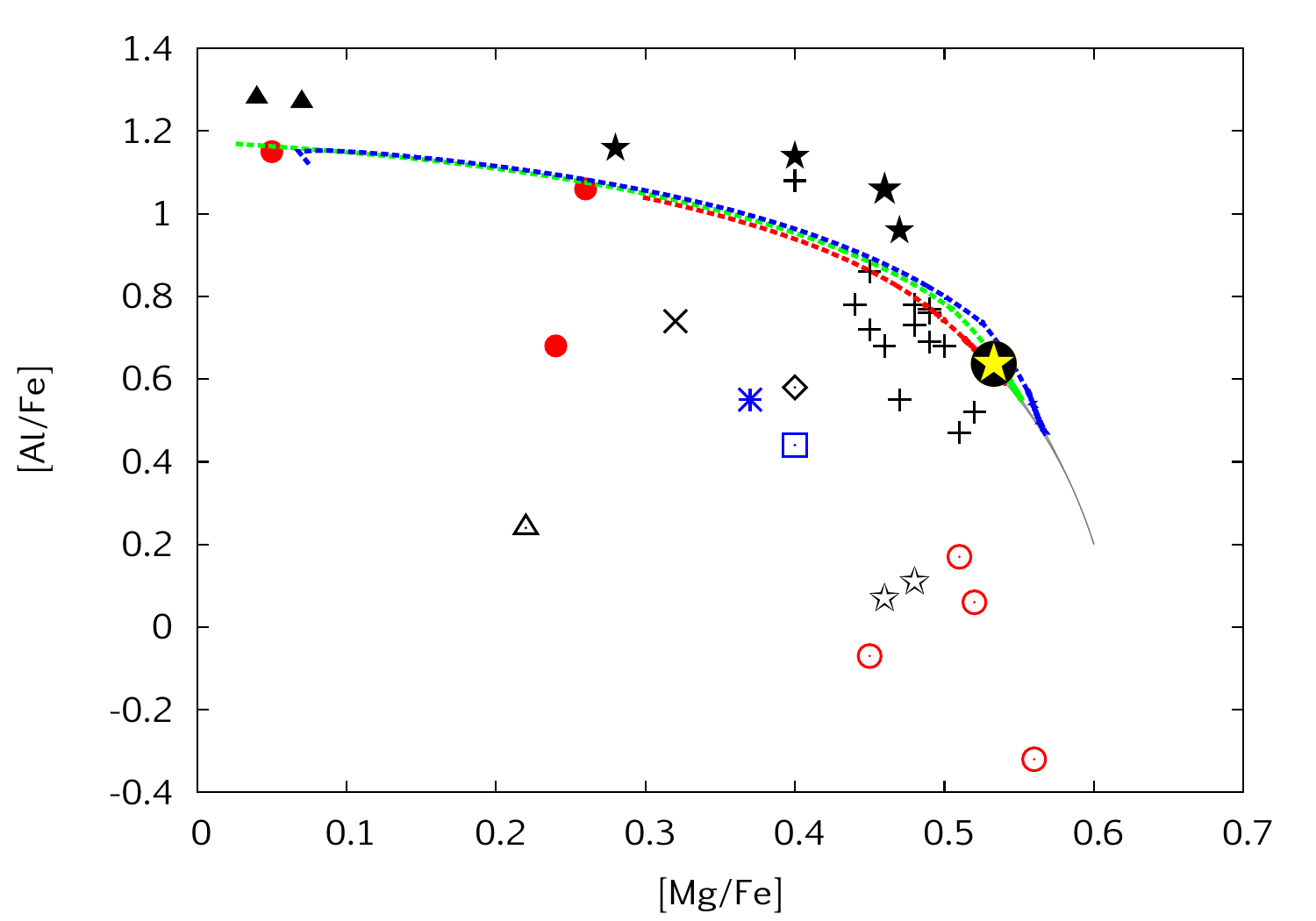}}
\caption{The same as Fig.~\ref{fig:obsNaO} but for Mg and Al.}
\label{fig:obsMgAl}
\end{figure}

\begin{figure}
\centering
\resizebox{0.8\hsize}{!}{\includegraphics[width=0.8\columnwidth,page=3]{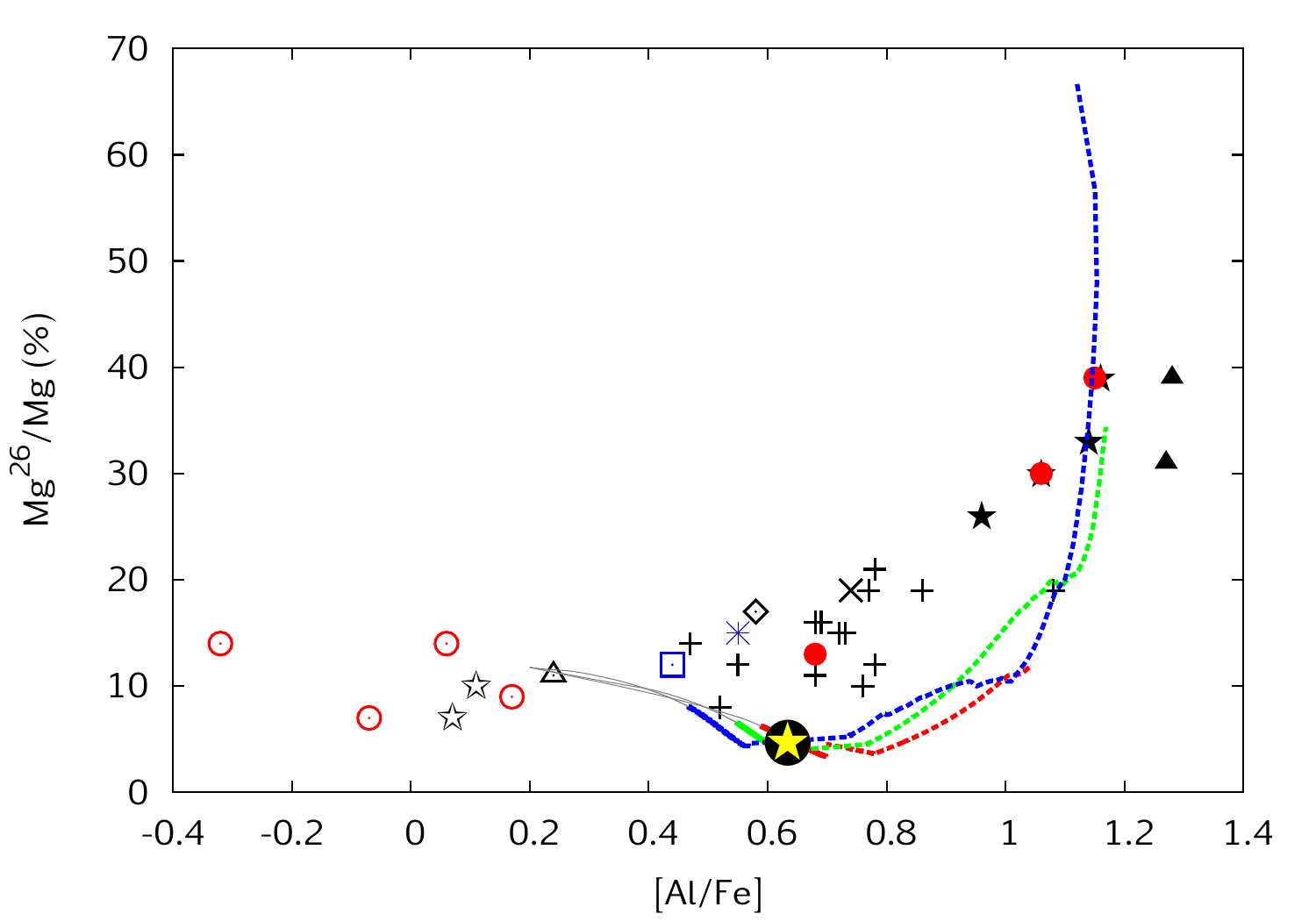}}
\resizebox{0.8\hsize}{!}{\includegraphics[width=0.8\columnwidth,page=4]{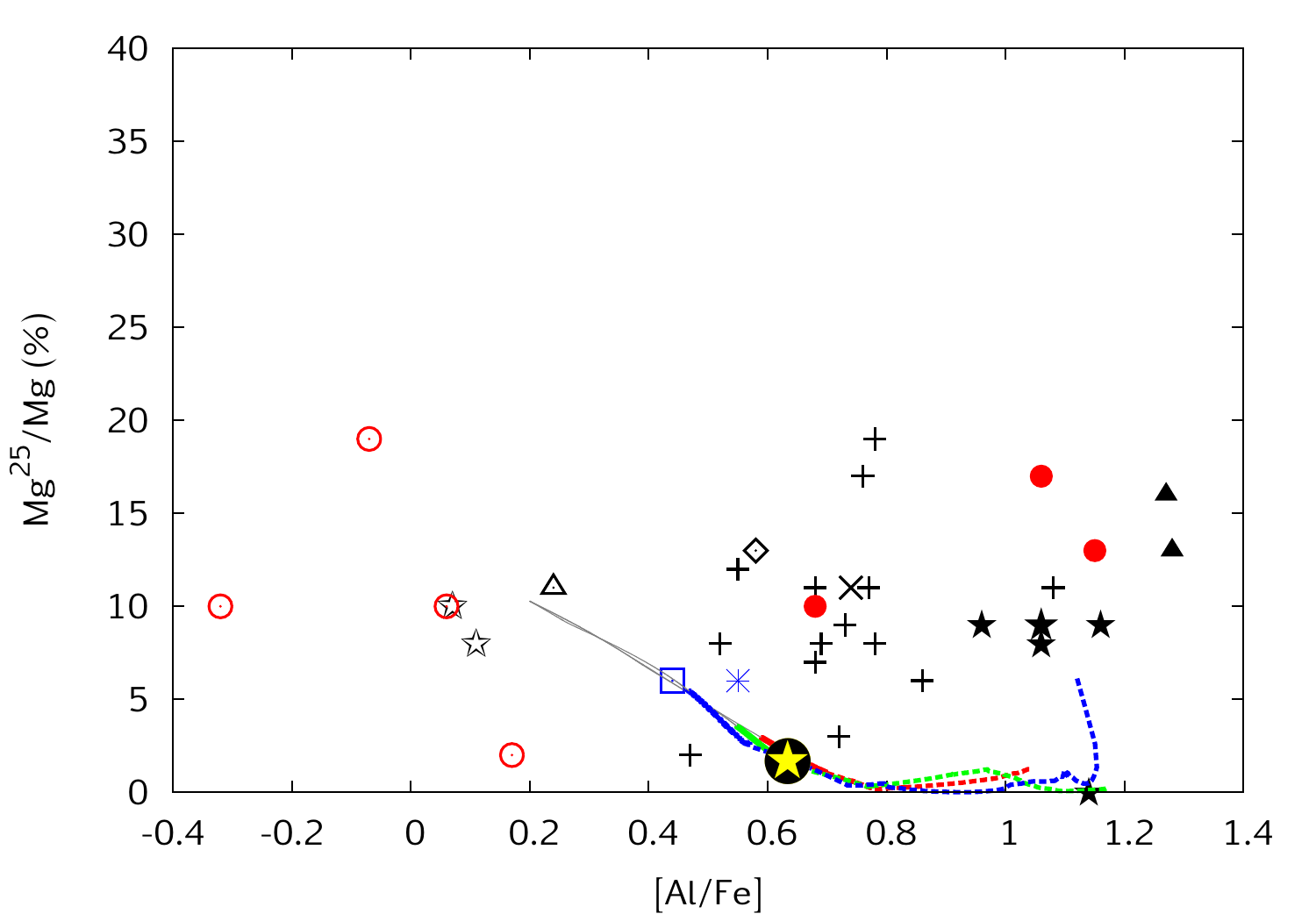}}
\resizebox{0.8\hsize}{!}{\includegraphics[width=0.8\columnwidth,page=5]{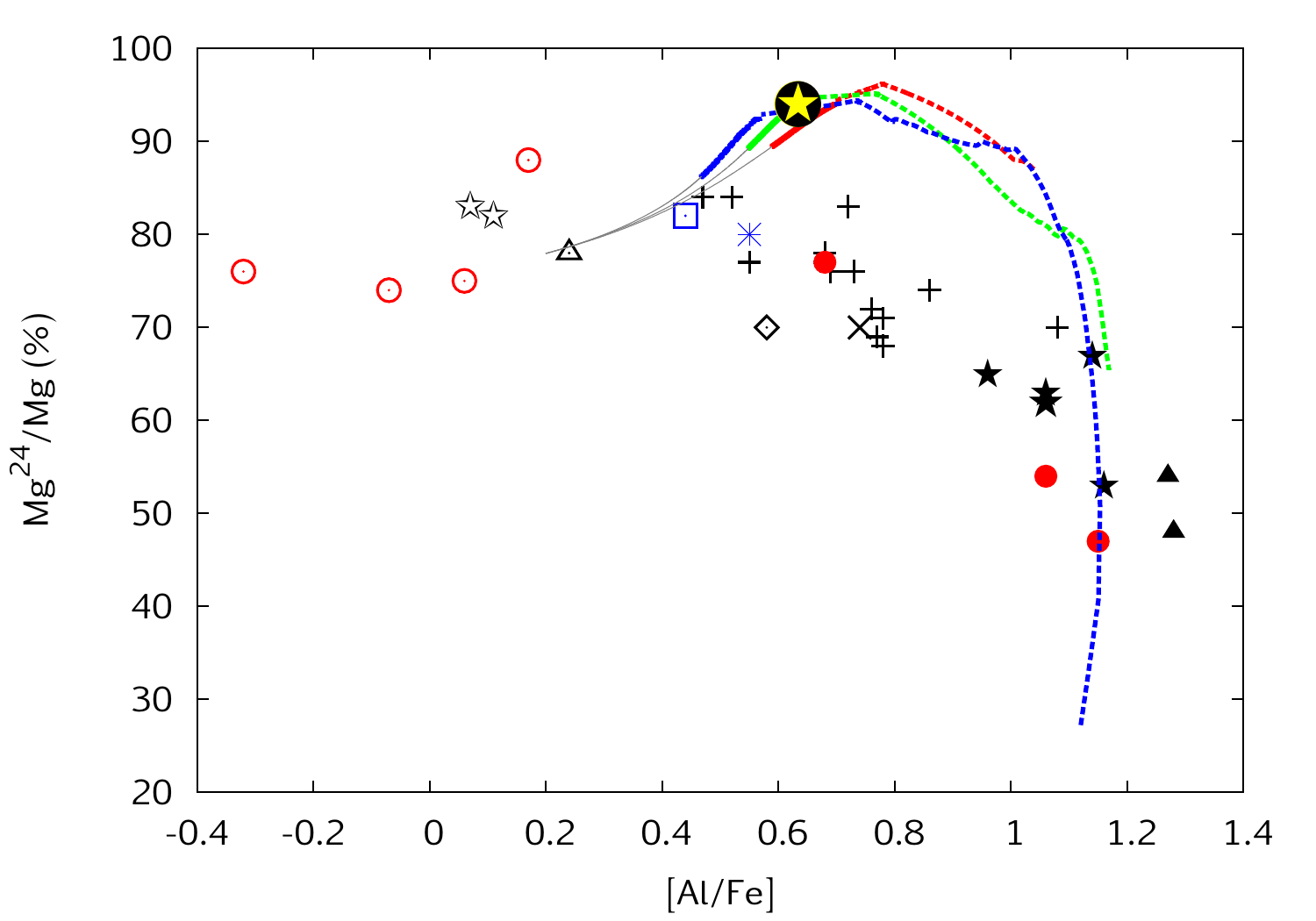}}
\caption{The same as Fig.~\ref{fig:obsMgAl} but for the isotopes of Mg.}
\label{fig:obsMgiso}
\end{figure}

The Mg-Al surface abundances of our models are shown in Fig.~\ref{fig:obsMgAl}. The surface Mg and Al abundances cover only a small fraction of all the observed variations in these elements. However, losing deeper layers of the envelope could explain the whole observed ranges of Mg and Al variations. When it comes to Mg, it is not only the sum of all three Mg-isotopes that is measured, but the ratios of them as well \citep{Yong:2003,Yong:2006,DaCosta:2013}. Fig.~\ref{fig:obsMgiso} shows the observed isotopic ratios of Mg as a function of the Al-abundance. As mentioned above, our models can reproduce the most extreme Al-abundance values observed in the case where deeper layers of the models are lost. In these deep layers, the Mg-isotopes also follow the observed trend: $^{24}$Mg is decreasing, $^{25}$Mg is slightly decreasing and $^{26}$Mg is considerably increasing compared to their values at the surface.

Due to the high core temperatures, the Mg-Al chain is very effective in our cool SG models. This is a clear advantage of our scenario: for example, neither the fast rotating star scenario nor the massive binary scenario can reach the required spread in Al and Mg, or reproduce the extreme ratios of the Mg-isotopes, unless the reaction rate of the Mg-Al chain is artificially increased \citep{Decressin:2007,deMink:2009}. 

From the comparison of our models' composition with the observed light-element abundances, we conclude that cool SG stars are promising candidates for the astrophysical source that pollutes the second generation of GC stars. 
Their strong, slow winds can enrich the interstellar material of the cluster with hot-hydrogen-burning products; the light-element abundances in their envelopes correspond to the most extreme pollution observed. If the stellar wind mixes with the pristine gas of the cluster \citep[as assumed for all other scenarios, such as the asymptotic giant branch star, the fast rotating star and the massive binary scenarios,][]{Bastian:2015}, this mixture can form stars with all of  the observed abundance spreads. Thus, cool SGs should be considered as potential contributors of the general pollution of GCs.

However, here we discuss our cool SG models' role not in the general pollution of the interstellar medium of GCs, but in the context of another star-forming process: low-mass star formation in a photoionization-confined shell around the cool SGs. To predict the composition of the SG-ejecta and thus the composition of the second generation of low-mass stars, we need to sum over the surface composition of the SG evolutionary models. We come back to this issue in Sect.~\ref{sec:comp}. In the following, we introduce the concept of the star-forming SG shell.


\section{Starformation in the shell}\label{sec:SGshells}

\subsection{Conditions in young GCs}\label{sec:Conditions}

Apart from the core-hydrogen-burning cool SGs, another important prediction by \citet{Szecsi:2015} is that the fast rotating massive stars become hot, compact and bright for their whole lifetime. These objects, called Transparent Wind UV-Intense (TWUIN) stars, have similar surface properties to those of Wolf--Rayet stars, but differ in that their stellar winds are optically thin \citep[see also][for further discussions of these objects]{Szecsi:2015b,Szecsi:2017h}. TWUIN stars produce a huge amount of ionizing radiation during their lifetimes. According to \citet{Szecsi:2015}, TWUIN stars have a Lyman-continuum luminosity of $Q_0\approx 10^{50}-10^{51}$~s$^{-1}$. A SG located 0.5~pc from such a star is therefore exposed to an ionizing photon flux, $F_{\gamma}$, between $3.3\times 10^{12}$~cm$^{-2}$~s$^{-1}$ and $3.3\times 10^{13}$~cm$^{-2}$~s$^{-1}$. In a dense cluster it is possible for the separation to be even smaller, leading to potentially even more extreme irradiating fluxes.

Following \citet{Szecsi:2015}, we suppose that $\sim$20\% of all massive stars rotate faster than required for quasi-homogeneous evolution, i.e.\ TWUIN-star formation. \citep[This ratio is supported by the rotational velocity distribution of massive stars in the Small Magellanic Cloud observed by][]{Mokiem:2006}. Thus, we have a population of massive stars in a young globular cluster where $\sim$80\% of stars evolve towards the supergiant branch while $\sim$20\% stay hot and emit ionizing radiation.

Supposing that the ionizing-radiation field of the TWUIN stars is isotropic, the wind structure of the SG stars changes significantly: their winds are photoionized from the outside in.  At the interface between ionized and neutral material, a dense, spherical shell developes, if the wind is sufficiently slow. This region is called the photoionization-confined shell.

\subsection{Photoionizaton-confined shells around cool SGs}\label{sec:pico}

\citet{Mackey:2014} developed the photoionization-confined shell model to explain the static shell observed around Betelgeuse, a nearby red SG star. According to their calculations, pressure from the photoionized wind generates a standing shock in the neutral part of the wind and forms an almost static, photoionization-confined shell. The shell traps up to 35\% of all mass lost during the red SG phase, confining this gas close to the central object until its final supernova explosion. 

We carried out simulations of a shell around a low-Z very massive SG star that undergoes core hydrogen burning. We use the \textsc{PION} code with spherical symmetry \citep{Mackey:2012} to simulate an evolving stellar wind that is photoionized by external radiation. The source of the ionizing radiation are the fast-rotating TWUIN stars, creating an isotopic radiation field that surrounds the SG star. The simulations follow \citet{Mackey:2014} except that we include stellar evolution and we use non-equilibrium heating and cooling rates for the gas thermal physics \citep[as in][]{Mackey:2015}. The stellar wind flows through the inner boundary of the grid with properties taken from the model with M$_{\rm ini}$=257~M$_{\odot}$ of \citet[][also see Sect.~\ref{sec:evolution}]{Szecsi:2015}. This evolutionary model has an initial rotational velocity of 100~km~s$^{-1}$ and mass loss in the SG phase of about 10$^{-3.5}$~M$_{\odot}$~yr$^{-1}$. 

The wind is initially cold (200~K; this has no effect on the results because the wind is highly supersonic). The wind velocity is calculated from the escape velocity following \citet{Eldridge:2006}, except that we set the SG wind velocity to be $v_\infty=0.1v_\mathrm{esc}$ for $T_\mathrm{\rm eff}<4500$ K. 
The above modification gives a minimum value of $v_\infty\approx12$~km\,s$^{-1}$. The simulations are run with a total metallicity of 0.0002 and surface abundance mass fractions X=0.5 and Y=0.4998, similar to the surface abundances in the low-Z stellar model \citep{Szecsi:2015}. 
The wind is exposed to an ionizing photon flux of $F_{\gamma} = 10^{13}$~cm$^{-2}$~s$^{-1}$ (cf.\ Sect.~\ref{sec:Conditions}) in the calculations presented here.

The formation of the shell in the simulation depends on the thermal physics of the shocked wind (which must be able to cool into a dense and cold layer); this is rather uncertain because we have no constraints on dust formation in such low-metallicity SGs. We use atomic line cooling \citep{Wolfire:2003} as implemented in \citet{Mackey:2013}, scaled to the metallicity of the stellar wind.

\begin{figure}
\centering
\resizebox{\fs\hsize}{!}{\includegraphics{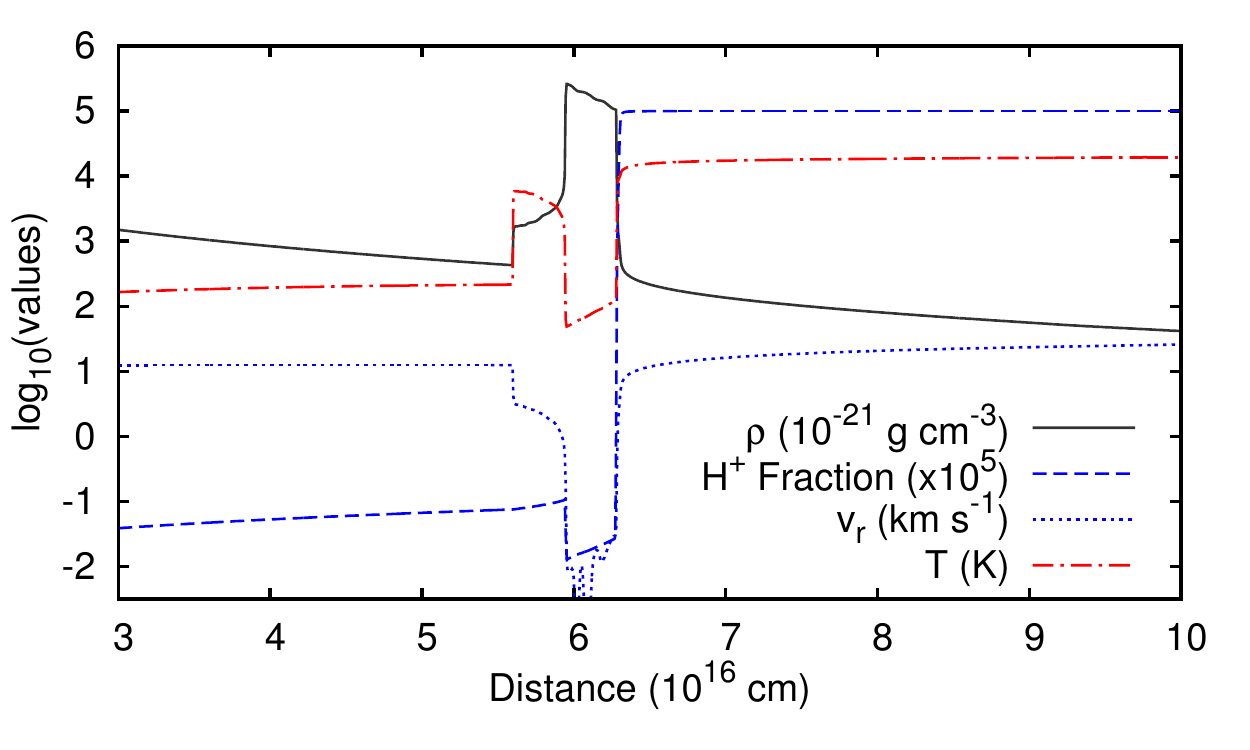}}
\caption{
Density, temperature, velocity, and ionization fraction for the simulation of the photoionizaton-confined shell around a core hydrogen burning supergiant with initial mass of 257~M$_{\odot}$. The snapshot is taken at the end of the stellar evolution calculation, when the star has an age of 1.88~Myr, at which time the shell mass is 14~M$_{\odot}$.
}
\label{fig:shelldens}
\end{figure}

\begin{figure}
\centering
\resizebox{\fs\hsize}{!}{\includegraphics{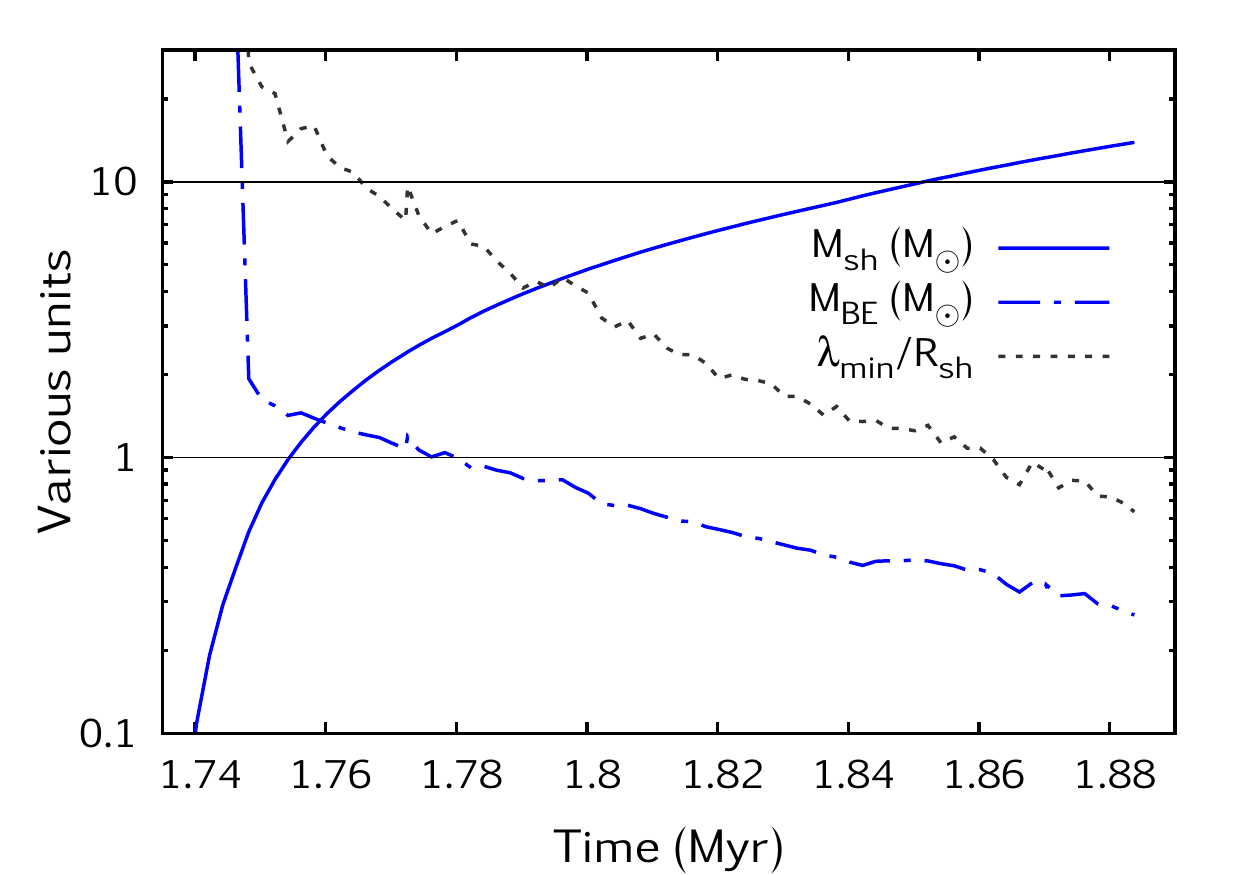}}
\caption{
Shell mass, $M_\mathrm{sh}$, as a function of time since the star's birth (solid blue line), compared to the Bonner-Ebert mass $M_\mathrm{BE}$ at the densest point in the shell (dot-dashed blue line). The dashed black line shows the minimum unstable wavelength in units of the shell radius.
}
\label{fig:shellmass}
\end{figure}

Fig.~\ref{fig:shelldens} shows the structure of the shell. The shell formed at a radius $r\approx0.02$~pc ($6\cdot10^{16}$ cm) from the supergiant (recall that the radius of the stellar model itself is 3.4$\cdot$10$^{14}$~cm, see Sect.~\ref{sec:evolution})
and shows the classic structure of a radiative shock: 
(i) an initial density jump at the shock of a factor of $\approx4$ with associated jumps in temperature and velocity according to the Rankine-Hugoniot jump conditions;
(ii) a cooling region where the temperature decreases with $r$, the density increases, and the velocity decreases; and
(iii) a cold dense layer.
The cold layer is bounded on the outside by the ionization front, at which radius the stellar wind is heated to $\approx12\,000$ K.
A thermally driven wind accelerates outwards from the ionization front.
We find that at the metallicity of the SG, the atomic cooling simulation produces a shell with density $\rho\approx2\times10^{-16}$~g\,cm$^{-3}$ and temperature $T\approx50$ K.

The shell mass, $M_{\rm shell}$, is plotted as a function of time in Fig.~\ref{fig:shellmass}.
It grows to M$_{\rm shell}\approx$~14~M$_{\odot}$ by the end of the simulation.
The Bonner-Ebert mass (i.e.\ the mass limit of the overdense region, above which the material collapses into a proto-star), $M_\mathrm{BE}$, and the minimum unstable wavelength $\lambda_\mathrm{min}$ are also plotted in Fig.~\ref{fig:shellmass}. They are discussed in the next section.

\subsection{Gravitational instability in the shell}\label{sec:grav}

For the second generation of low mass stars to form in the photoionization-confined shell, the shell should be gravitationally unstable. To show that the shell in our simulation is indeed gravitationally unstable against perturbations, we follow \citet[][see their eqs.~2.12-2.14]{Elmegreen:1998} who describes the stability of a shocked sheet of gas \citep[see also][]{Doroshkevich:1980,Vishniac:1983}. The dispersion relation (eq.~2.13) gives the condition that perturbations with wavelength~$\lambda$ are unstable~($\omega^2>0$)~if
\begin{equation}
\lambda \geq \frac{c^2}{G\sigma} = \frac{P}{G\sigma\rho} 
\end{equation}\label{eq:lambda} 
where $c$ is the isothermal sound speed defined by $c^2\equiv P/\rho$ ($P$ being the thermal pressure and $\rho$ the density), and $\sigma$ is the column density through the shell. This condition needs to be fulfilled by the shell in order to become gravitationally unstable. We define $\lambda_\mathrm{min}$ to be the wavelength at which this inequality is an equality.

In our simulation, the shell thickness is \mbox{$l=0.36\cdot10^{16}$~cm}, density is \mbox{$\rho=2.65\cdot10^{-16}$~g~cm$^{-3}$}, and  pressure is \mbox{$P=5.89\cdot10^{-7}$~dyne~cm$^{-2}$}. For this shell, the above condition gives a perturbation wavelength \mbox{$\lambda_\mathrm{min} = 3.4\cdot10^{16}$~cm}. 

An overdense region should have a diameter of $\lambda$/2. For our spherical shells, we should restrict $\lambda$/2 to be significantly less than the radius of curvature, so that the unstable part of the shell looks more like a flat sheet than a sphere. The shell is at radius \mbox{$\sim$6.2$\cdot10^{16}$~cm} (0.02~pc). The angular size of the overdense region is thus
\mbox{$\lambda_\mathrm{min}/2R_\mathrm{sh}=1.7/6/2\approx0.3$} which is much less than one radian (about 16\degr), so curvature effects are relatively small.
Fig.~\ref{fig:shellmass} shows that \mbox{$\lambda_\mathrm{min}/2R_\mathrm{sh}\approx0.33$} at the end of the simulation, similar to the estimate above.

The Bonnor-Ebert mass for this dense region is
\begin{equation}
M_{\rm BE}=1.18\frac{c^4}{P^{1/2}G^{3/2}}=0.2~ {\rm M}_{\odot},
\label{eq:BEmass}
\end{equation}
meaning that if the dense region contains more mass than this, it would collapse to a protostar. The mass of the dense region depends on its geometry, but with a density of \mbox{$\rho=2.65\cdot10^{-16}$~g~cm$^{-3}$} and a length scale of \mbox{$\lambda/2 \approx 1.7\cdot10^{16}$~cm}, it is around 2-3~M$_{\odot}$.
We see from Fig.~\ref{fig:shellmass} that the shell contains a mass $M_\mathrm{sh}\approx50M_\mathrm{BE}$ at the end of the simulation. 

The stability analysis shows that the shell does not become unstable until it contains $\geq20M_\mathrm{BE}$ because the mass is distributed in a shell and not in a spherical cloud. We conclude therefore, that the thermodynamic conditions in the shell allow for gravitational instability, and that potentially many low mass stars may form from a single shell.

\subsection{Forming the second generation of stars in the shell}

Even if gravitational instabilities develop in the shell, the protostars should have been formed before the shell evaporates. This means that the growth timescale of the perturbation should be less than a few times 10$^5$ years (cf.\ lifetimes of SG stars in our simulation, Sect.~\ref{sec:evolution}). Using eqs.~2.12 and 2.14 from \citet{Elmegreen:1998}, we get 3100 and 2.2$\cdot$10$^4$ years, respectively. These timescales are indeed significantly shorter than the life of the SG star with shell.

Once gravitational instability sets in, the collapse timescale is very short because the shell already has a very high density, much larger than dense cores in molecular clouds. Three-dimensional simulations are required to follow the gravitational collapse, so we cannot predict the final masses of the stars that will form. They may be larger than $M_\mathrm{BE}$ because the shell is constantly replenished from the cool SG's mass-loss, and this could accrete onto collapsing cores. 

It is highly unlikely, however, that this star-formation channel would have a typical initial mass function. It will rather be dominated by stars with less than one solar mass, and the probability of forming massive stars is expected to be extremely small. On the other hand, we also do not expect very low-mass stars since our simulation predicts a typical mass of 0.2~M$_{\odot}$ for proto-stars, and they are probably still accreting.

Star formation could be a bursty process if gravitational instability sets in at the same time everywhere in the shell (i.e.~if the shell is homogeneous), or more continuous if the shell is asymmetric and/or clumpy.
In either case, star formation does not destroy the shell, but rather makes space for further gas accumulation and subsequent collapse to form more stars. 
After the shell begins to collapse, its gaseous mass (excluding protostars) is determined by the addition of new material from the stellar wind of the cool SG, balanced by the collapse of shell material to form new stars, plus accretion of shell material onto existing protostars.
The addition of new material is about 35\% of the cool SG's mass-loss rate, so $\sim10^{-4}$~M$_\odot$~yr$^{-1}$.
Accretion rates onto low-mass protostars are typically $10^{-7}$~M$_\odot$~yr$^{-1}$ \citep{Hartmann:1996}, and so this is unlikely to affect the shell mass because the shell can only form $\approx 10-50$ protostars at any one time (recall, it becomes unstable when its mass is $\gtrsim 10$~M$_\odot$). The reservoir of gas available to form new stars is therefore determined by the mass-loss rate of the cool SG and the rate at which new protostars are condensing out of the shell.

This means that star formation in the shell is expected to be a more or less continuous (but stochastic) process. After the shell has formed and grown to become unstable, some bits of it collapse at different times. But in the meantime, the shell-material is constantly replenished by the SG wind. Thus, an equilibrium develops between mass added to the shell and mass lost through star formation.

\subsection{Composition of the stars in the shell}\label{sec:comp}

The low-mass stars formed in the shell necessarily reflect the composition of the SG wind which is polluted by hot-hydrogen-burning products. 
To compute the composition of the shell-stars, we assume that the wind that leaves the SG star goes directly into the shell, and that the material inside the shell is homogeneously mixed. We take into account that the shell only traps a certain amount of the wind-mass (as follows from the hydrodynamical simulations of its structure presented in Fig.~\ref{fig:shellmass}), and thus sum over the wind composition. 

Figs.~\ref{fig:obsNaO} and \ref{fig:obsMgAl} show the composition of a star formed inside the shell simulated around the M$_{\rm ini}$=257~M$_{\odot}$ supergiant. The abundances of Na and O of the shell-stars are compatible with the surface composition observed in the extremely polluted population.
The abundances of Mg and Al of our shell stars are compatible with the intermediate population. To fit more extreme abundances of Mg and Al, deeper layers of the SG star should be lost (represented by the dashed lines in Fig.~\ref{fig:obsMgAl}). This could still happen during the post-main-sequence evolution of the SG model which would last for an additional 0.17~Myr (not simulated). 
The shell stars have a helium mass fraction of Y$_{\rm sh}$=0.48. We discuss the issue of the observed helium abundance of GC stars in Sect.~\ref{sec:helium}. 

\section{Discussion}\label{sec:discuu}

\subsection{Mass budget}\label{sec:massb}

Any scenarios that aim to explain the abundance anomalies observed in GCs need to account for the mass that is contained in the first as well as in the second generation of stars. The three most popular of the polluter sources (asymptotic giant branch stars, fast rotating stars, massive binaries), when only one of them is taken into consideration, fail to explain the amount of stellar mass that we observe with polluted composition. These scenarios suppose that the polluted material stays inside the gravitational potential well of the cluster, preferably accumulating near the center. There the polluted material mixes with the pristine material and forms the second stellar generation. This would explain why we observe not just the primordial and extreme abundances but everything in between (see the observations in Figs.~\ref{fig:obsNaO}~and~\ref{fig:obsMgAl}). But for a second generation to be as numerous as the first generation, one needs much more polluted material than one of these sources can provide \citep{deMink:2009}. Therefore, it is possible that more then one pollution source is present, or even that all the suggested sources contribute \citep{Bastian:2013}.

The mass budget constraint in its simplest form is the following: the second generation that is born inside the shell should contain as much (50:50) mass as the first generation of low-mass stars born normally. (The ratio 50:50 is applicable for the GCs with average mass, but there is evidence that higher-mass clusters have a higher fraction of second generation stars, see Sect.~\ref{sec:verymassive}.) 

\subsubsection{Classical IMF}\label{sec:Salpeter}

To investigate the mass budget of our starforming shell scenario, we follow the discussion of \citet{deMink:2009}. Namely, we apply an initial mass function (IMF) between 0.1-1000~M$_{\odot}$ to represent the first generation of stars, as follows \citep{Salpeter:1955,Kroupa:2001}:
\begin{equation}
\begin{split}
N(m)=
\begin{cases}
0.29\cdot m^{-1.3}, & \text{if}\ \ 0.1<m<0.5 \\
0.14\cdot m^{-2.3}, & \text{if}\ \ 0.5<m<1000
\end{cases}
\end{split}\label{eq:imf}
\end{equation}

We take the low-mass stars in the first, unpolluted generation to be between 0.1-0.8~M$_{\odot}$, that is, the mass of stars observed in GCs today \citep[see][]{deMink:2009}.
As for the shell-forming SGs in the first generation, we argue that our models are representative for them in the mass range of 80-1000~M$_{\odot}$. This argument is justified because (1) mass-loss rates in this mass range are high enough for massive shells to form (cf. Sect.~\ref{sec:lowmass}) and because (2) models in this mass range are expected to become core-hydrogen-burning SG stars (cf. Sect.~\ref{sec:evolution}). Additionally, we assume here that the second generation of shell-stars also form between 0.1$-$0.8~M$_{\odot}$, following the mass-distribution of the unpolluted first generation of low-mass stars. We discuss the consequences of \textit{not} assuming this in Sect.~\ref{sec:2GIMF}.

Eq.~\ref{eq:imf} predicts that the first generation of low-mass stars represent 35\% of the total stellar mass initially present in the cluster. Thus to fulfil the mass budget constraint, the second generation should also account for the same, 35\% of the total mass. 
Unfortunately, the mass of the SG stars represent only 10\% of the total. If it would be lost through the wind and incorporated into the second generation in the shell with an efficiency of $\xi=$100\% (which is clearly a very weak constraint not only because it would require an unreasonably high mass-loss rate but also because we expect $\sim$20\% of all massive stars to be hot TWUIN stars, see Sect.~\ref{sec:Conditions}), this is still far from the 35\% we aim to account for. 

\subsubsection{Top-heavy IMF}\label{sec:topheavy}

One simple way around this issue is to assume a top-heavy IMF, which has indeed been favoured for massive clusters recently \citep{Ciardi:2003,Dabringhausen:2009}.
For example, \citet{Decressin:2010} suggests a flat IMF with index $-$1.55 (instead of $-$2.3 as in~Eq.~(\ref{eq:imf})) to make their fast rotating star scenario work. Our SG shell scenario, however, can work with less extreme values.
Assuming that the massive component of the IMF has an index of $-$2.07 (instead of $-$2.3), the first generation low-mass stars (0.1-0.8~M$_{\odot}$) represent 23\% of the stellar mass initially present in the cluster, while the SG stars (80-1000~M$_{\odot}$) also represent 23\%, satisfying the weak constraint mentioned at the end of Sect.~\ref{sec:Salpeter}.

A strong constraint should take into account: (1) that only $\sim$40\% of the SG mass is lost in the wind; (2) that the shell contains only $\sim$35\% of the wind mass; and (3) that only $\sim$80\% of massive stars evolve towards the supergiant branch (the rest are the TWUIN stars responsible for the ionization). Thus, the mass contained in SG stars will be converted into low-mass stars with an efficiency of $\xi$~$\approx$~40\%~$\times$~35\%~$\times$~80\%~$\approx$~12\%. With this efficiency, an IMF index of $-$1.71 is needed, which translates to 7\% of the total mass in first generation low-mass stars (i.e.\ 0.1-0.8~M$_{\odot}$), and 55\% of the total mass in massive stars (i.e.\ 80-1000~M$_{\odot}$). The mass budget problem is then solved because from this 55\%, only 55\%~$\times$~$\xi$~$\approx$~7\% will be converted into the second generation of low-mass stars.

However, we may not need this strong constraint, since the ratio of the material trapped in the SG shell should be higher than 35\%, which is the nominal value in our simulation. Thus the efficiency, $\xi$, of converting SG mass into shell-stars may be significantly higher than 12\%. The reason for this is that, according to the speculation at the end of Sect.~\ref{sec:grav}, the shell may retain more wind material than the nominal value since the proto-stars are constantly accreting. Since accretion is not included into our shell-simulation, we cannot properly quantify that at this point. Nonetheless, the weak and the strong constraints presented above correspond to IMF indices of $-$2.07 and $-$1.71, respectively, so we conclude that the index required for our scenario to work should be somewhere between these two values.

\subsubsection{On the number of stars in the cluster and in the shell}\label{sec:number}

We give an order of magnitude estimate of the number of stars present in a typical GC where SG shells are forming the second generation. To do this, we assume an average GC with total mass of 10$^5$~M$_{\odot}$ and with an IMF index $-$1.71. This IMF allocates 7\% of the total mass into first generation stars between 0.1-0.8~M$_{\odot}$, and 55\% into SG stars between 80-1000~M$_{\odot}$ (while the rest has no mass-contribution to this particular scenario). The mass of stars in the second generation (i.e.\ formed from shells around SGs) also represents 7\%.

We take 257~M$_{\odot}$ to be the representative mass for the massive regime (that is, the initial mass of the SG model around which our simulation was carried out); and we take 
0.2~M$_{\odot}$ to be the representative average mass for both the first generation low-mass stars and the second generation of shell-stars. This value, 0.2~M$_{\odot}$, is the Bonnor-Ebert mass of the objects in our simulation presented in Sect.~\ref{sec:grav}, so it may depend on the mass and geometry of the shell and, therefore, on the mass of the SG.

With these assumptions, the first generation of low-mass stars consist of 35\,000 stars, and so does the second generation. Besides, the first generation must have contained 214 stars in the massive regime. From these, 171 should evolve to be supergiants and have shells, and 43 should be hot TWUIN stars. Note however that there are much more ionizing sources than that, since fast rotating models in the mass range of 9$-$80~M$_{\odot}$ also predict TWUIN stars \citep{Szecsi:2015}. 

To form 35\,000 second-generation stars, all 171 supergiants have to form $\sim$200 low-mass stars of 0.2~M$_{\odot}$ out of its wind material. One may recall from Sect.~\ref{sec:grav} that the structure of our simulated shell facilitates the formation of only 50 protostars of this mass at any given time, and that the protostars condensing out of the shell make space for further gas accumulation and subsequent collapses. Thus, from the mass budget constraints it follows that the shell in our simulation should undergo $\sim$3-4 subsequent events of gravitational collapse. 

We say subsequent collapses, but we are not suggesting that the shell will form, then everywhere collapse into stars, then re-form and re-collapse, and repeat again. What we suggest is that the shell will form, grow to become unstable, and then there will be stars forming out of cloud material \textit{all the time}. We do not expect it to be an episodic process, but rather a continuous one, resulting in $\sim$3-4 times 50 protostars at the end. 

\subsection{On very massive stars and very massive globular clusters}\label{sec:verymassive}

A crucial assumption of the star-forming-shell scenario is the presence of very massive stars in the young cluster. Very massive (>100~M$_{\odot}$) stars are theorized to form either via accretion (i.e.\ the same process that creates lower mass stars) or collision \citep[in extremely dense regions,][]{Krumholz:2014}. Therefore, it is not unreasonable to hypothesize stars as massive as this born in the young GCs. For example, \citet{Denissenkov:2014} assumed stars with 10$^4$~M$_{\odot}$ to give a possible explanation for the GC abundance anomalies.

Statistically, to find very massive stars in a star-forming region in significant number, either the mass of the region has to be large or the IMF has to be very top-heavy---or both. In Sect.~\ref{sec:number} we apply a top-heavy IMF of index $-$1.71 (coming from the strong constraint presented in Sect.~\ref{sec:massb}) and an average GC mass of 10$^5$~M$_{\odot}$ (which results in 171 SGs of the nominal mass 257~M$_{\odot}$). However, some GCs are significantly more massive than that. For example, the mass of $\omega$~Cen is 4$\cdot$10$^6$~M$_{\odot}$. 

It has been suggested that the fraction of enriched stars (and in general, the complexity of the multiple population phenomenon) correlates with cluster mass \citep{Carretta:2010,Piotto:2015,Milone:2017}. To account for this, we computed the IMF index not only for a 50:50 ratio of second vs. first generation, but also for a 70:30 ratio (as in some high mass clusters) and a for a 90:10 ratio (as in the highest mass clusters such as e.g. NGC~2808). In the case of a 70:30 ratio, an IMF index of $-$1.6 is needed to fulfill the strong constraint in our starforming shell scenario; while in the case of a 90:10 ratio, $-$1.4 is needed. So we conclude that if---for some reason---the IMF gets more top-heavy with cluster mass, our scenario may work to explain even the most massive clusters. But this argument also applies to all other self-enrichment scenarios involving massive stars, so it is not a distinguishing feature of our scenario. 

It is so far unclear if the same mechanism forms all galactic GCs. There is evidence that the low-metallicity GCs in the outer halo have been accreted from neighbouring dwarf galaxies, while the high-metallicity GCs in the inner halo have been formed in situ \citep{Brodie:2006,Forbes:2010}. Some of the most massive GCs, $\omega$~Cen amongst them, possibly used to be dwarf galaxies \citep{Schiavon:2017}. In short, the formation of globular clusters is a complex problem that may require several theoretical scenarios to work together; our scenario may be one of them. 

\subsection{Supergiants at lower masses}\label{sec:lowmass}

We presented SG models with initial masses between 150$-$575~M$_{\odot}$, and considered them representative for the mass range of 80$-$1000~M$_{\odot}$ when talking about the mass budget in Sect.~\ref{sec:massb}. The reasons for not including SG models with lower masses (9$-$80~M$_{\odot}$) into our analysis, are the following.

First, their mass-loss is too low to form shells around them. We recall from Sect.~\ref{sec:evolution} that the model around which we simulated the shell, has a mass-loss rate of $-$3.5~[log~M$_{\odot}$~yr$^{-1}$]. Our computations of SG models with 70, 43 and 26~M$_{\odot}$ show that they have mass-loss rates of $-$4.6, $-$5.2 and $-$5.9, respectively. The shells around them will not be massive enough for the second generation of stars to form: it takes a long time to build up a solar mass in the shell, let alone tens of solar masses, if log($\dot{\mathrm{M}}$)~$\sim$~$-$5.
The second problem is geometric. The shell will be closer to the star, and so have smaller volume and less physical space in which to grow. 

We cannot exlude, however, that the wind material of these lower-mass SG stars will be expelled into the cluster. There, it might be able to cool later on and -- possibly diluted with some pristine gas -- make new stars. Since these lower-mass SG stars are more likely to form, and thus would dominate over the very massive stars even with a top-heavy IMF, it is an important question to investigate their contribution to the cluster's chemical evolution. A detailed analysis of this scenario will be performed in another work. Our preliminary results nonetheless show that models below 80~M$_{\odot}$ evolve to the SG branch only during their core-helium burning phase. Their surface Na\&O composition reflects the primordial or intermediate population (as defined in Fig.~\ref{fig:obsNaO}), but not the extreme one. As for the Mg\&Al anticorrelation, they show some minor variation only in Al, but no variation in Mg. 

Recently, \citet{Schiavon:2017} implied that, at a fixed metallicity, some GCs show variation in Mg and some not. In particular, they detected 23 giant stars in some high-Z and low-Z GCs (situated in the inner Galaxy), and found no clear anti-correlation between Al and Mg. Instead, they report a substantial spread in the abundance of Al and a smaller spread in Mg; while they also admit that the their sample is too small for this to be statistically significant. Nonetheless, this is an interesting finding from our point of view, especially when we talk about lower-mass SGs with <~80~M$_{\odot}$. As we see only minor variation in Al and no variation in Mg, we speculate that---without quantifying their contribution at this point---the presence of SG stars with <~80~M$_{\odot}$ in young clusters may help us to explain why some GCs show variation in Mg and some not.

\subsection{Helium spread in different clusters}\label{sec:helium}

In some globular clusters, there are extremely helium-rich stars. For example, $\sim$15\% of the stars in NGC~2808 show helium abundance of Y$\sim$0.4, as inferred from their multiple main sequences \citep{Piotto:2007,DAntona:2007}, as well as from spectroscopic measurements \citep{Marino:2014}. Other GCs, however, have less extreme helium variations \citep[][]{Bastian:2015,Dotter:2015}. 

The most extreme values cannot be reproduced by asymptotic giant branch stars \citep{Karakas:2006}. All the other polluter sources (massive binaries, fast rotating stars, supermassive stars) have a general problem reproducing the required light element variations when the helium spread is a constraint, as shown by \citet{Bastian:2015}. 
The reason for this is that the Ne-Na and Mg-Al chains are side-processes of the CNO-cycle -- therefore, together with their burning products a significant amount of helium must be produced as well.

Our simulated shell-stars behave the same way as other massive polluters. Their surface composition (represented by the black-yellow symbol in Figs.~\ref{fig:obsNaO} and \ref{fig:obsMgAl}) contains helium: Y$_{\rm sh}$=0.48. Therefore, they can also only explain the pollution in Na-O and Mg-Al together with a high helium abundance, similar to other scenarios that involve massive stars. 

This issue is generic, as both the Ne-Na chain and the Mg-Al chain are side reactions of hot hydrogen-burning \citep{Bastian:2015,Lochhaas:2017}. Hydrogen burns into helium; therefore, whatever nuclear change occurs in the Na/O/Mg/Al abundances due to these chains, it will be accompanied by a change in helium abundance, unless we find a mechanism that separates Na/O/Mg/Al from helium either inside the star or in the interstellar material.

\subsection{Dynamical interactions}\label{sec:coll}

\subsubsection{Collisions of cluster members and shell}
Here we discuss issues about collisions of a random cluster star with a shell around a SG: how often may these collisions happen, and what consequences they may have.

Globular clusters have central densities $\gtrsim 10^3$~M$_\odot$~pc$^{-3}$ and typical stellar mass $0.8$~M$_\odot$ \citep{PortegiesZwart:2010} corresponding to a number density, $n_\star \gtrsim 10^3$~pc$^{-3}$.
They also have internal velocity dispersion, $\sigma_\mathrm{v}\approx 1-10$~km~s$^{-1}$ \citep{Harris:1996}.
The collision time, $t_\mathrm{coll}$, of one of their stars with a shell around a SG, with shell radius $R_\mathrm{sh}\approx0.02$~pc, 
can be derived from Eq.~(26) of \citet{PortegiesZwart:2010} as follows:

\begin{equation}
t_\mathrm{coll} \approx 0.16\;\mathrm{Myr}
\left(\frac{n_\star}{10^3\;\mathrm{pc}^{-3}}\right)^{-1}
\left(\frac{\sigma_\mathrm{v}}{5\;\mathrm{km~s}^{-1}}\right)^{-1}
\left(\frac{R_\mathrm{sh}}{0.02\;\mathrm{pc}}\right)^{-2}\label{eq:coll}
\end{equation}
According to this simple, order-of-magnitude estimate, on the order of one star will pass through the cool SG shell during its existence (it lasts $\sim10^5$~years).

However, central densities of \textsl{young} GCs might have been higher than assumed in Eq.~(\ref{eq:coll}). One argument for this is that YMCs, thought to be analogous to young GCs, have central densities much higher than observed today in GCs. A well-known example of a resolved YMC is the Arches cluster with central density of $10^5$~M$_\odot$~pc$^{-3}$ \citep{PortegiesZwart:2010}. Another argument is that the mass may be segregated (i.e.\ stars with masses greater than a given value are found to be more centrally concentrated than the average stellar mass) leading to a higher central density. Additionally, gravitational focusing (i.e.\ enhanced probability that two stars will collide due to their mutual gravitational attraction) may play a role.

If the central stellar density is higher than assumed in Eq.~(\ref{eq:coll}), this means two things. First, this would lead to more (potentially destructive) collisions. In such a dense environment as the Arches cluster, the estimated collision time is two orders of magnitudes higher than in Eq.~(\ref{eq:coll}). Second, the ionizing sources would be closer to the SGs if the central density is higher. Thus, the ionising flux would be larger and the shells would be more compact. This would decrease the probability of a collision, balancing the first effect. 

Whether the interaction with a star of low-mass would enhance or inhibit star formation in the SG shell is not clear, and would require complex simulations to model accurately.
If, on the other hand, the star were massive with a strong wind and large Lyman-continuum luminosity, then it would have a strong disruptive effect on the shell.
This may be happening to the wind of the red supergiant W26 in Westerlund~1
\citep{Mackey:2015}.
The probability of a massive star passing through a cool SG shell is small, however, because even the top-heavy mass function prefers low-mass objects (cf.\ the discussion on the number of stars in Sect.~\ref{sec:massb}).

Finally, we point out that even if the shells are destroyed by collision, their material may sink into the cluster core. It is possible that, independently of the formation of SG shells, the material in the cluster core is constantly forming stars, as supposed by many other scenarios \citep[cf.][]{Bastian:2015}. Our supergiants are therefore expected, even with their shells destroyed, to contribute to the chemical evolution of the young cluster by expelling polluted gas into the intracluster medium.

\subsubsection{The probability of falling into the SG}

Once the second generation of stars form in the shell, they are no longer subject to the radiation pressure from the central SG. The radial velocity of the shell-stars is therefore quite small. But it is not zero. In Galactic star formation, the clouds and the dense cores have velocity dispersions larger than the sound speed, attributed to supersonic turbulence \citep{MacLow:2004}. The shell around the SG will be the same, and so we expect that the dense cores that collapse to form stars will have non-radial velocities that are at least comparable to the local sound speed, and probably larger. 

While detailed star-formation simulations and N-body dynamics calculations would be required to address this problem, we can present a simple estimation here to demonstrate our point. For T~=~70~K, the sound speed is about 0.6~km~s$^{-1}$. For a 250~M$_{\odot}$ supergiant, and a shell at 6.0$\cdot$10$^{16}$~cm from the star, the escape velocity is 3.3~km~s$^{-1}$ and the circular velocity is 2.4~km~s$^{-1}$. This means that the random non-radial motions are, on average, >~25\% of the circular orbital velocity, and so the shell stars will be on elliptical orbits. 

The probability of actually falling into the supergiant is thus very small when simply considering orbits.  It is not obvious whether N-body interactions between the many protostars in the shell would eject stars into the cluster and/or increase the likelihood of collision with the central supergiant, and we cannot make predictions at this stage.

\subsection{On high-metallicity clusters and future plans}

Our work focuses on low-metallicity since the majority of GCs with abundance anomalies are between [Fe/H]=$-$2.0 and $-$1.0. We suspect that our model of star-formation in shells will hardly work at high-metallicity. As shown by the models of \citet{Brott:2011} and \citet{Koehler:2015}, massive stars with LMC metallicity do indeed experience envelope inflation above 40~M$_{\odot}$. However, the very massive ones ($\gtrsim$150~M$_{\odot}$) do not become cool supergiants because their mass-loss is very high and so they become hot Wolf--Rayet stars instead. We do not expect shells to form around these hot stars. As for the LMC models between 40$-$150~M$_{\odot}$, they do evolve to the supergiants branch. So they may form shells, although it is beyond the scope of the present work to simulate such a shell and analyse its stability.

The fact that multiple populations have not been found in nearby super star clusters to date \citep{Mucciarelli:2014} may mean, in the context of our scenario, that either (1) SG shells are not stable at high-metallicity, or (2) they do not create (too many) new stars, or even (3) that the composition of the new stars is indistinguishable from that of the old ones. Indeed, our preliminary investigation of the LMC models shows that they have lower core temperatures than their low-Z counterparts, and so the Mg-Al~chain is not effective in them. Thus, even if the second generation is formed in a high-metallicity cluster, we do not expect them to show significant Mg/Al variations. As for the other elements, the variations of Na/O in the winds of the LMC models are present, but more moderate than in our low-metallicity models.

On the other hand, some of the higher-metallicity GCs also have multiple populations \citep[as observed by e.g.][]{Schiavon:2017}. This however does not mean that the same scenario produces the multiple populations at all metallicities. As mentioned in Sect.~\ref{sec:verymassive}, we do not expect that the complex problem of GC formation would be solved by one simple scenario. Indeed, both our low-Z models and the LMC models can be applied in another scenario, in which the mass lost in winds from massive stars can later cool in the cluster core and form new stars (cf.~Sect.~\ref{sec:lowmass}).

Detailed investigation of both sets of models and their wind composition, as well as the possible ways their strong wind may influence the chemical and hydrodynamical evolution of their clusters, are planned in the future.

Indeed, the metallicity-dependence of our scenario, along with that of other scenarios in the literature, should be investigated. Some observations \citep[such as the compilation of photometric results from the
HST~UV~survey by][which is mainly tracing N-abundance variations]{Milone:2017} imply that there is no clear relation between the fraction of stars in each population and the metallicity
of the host cluster. From the modelling point of view, we can say the following about metallicities between our models and the LMC models. \citet{Sanyal:2017} showed that we can expect core-hydrogen-burning SG stars with the composition of the Small Magellanic Cloud (SMC) with luminosities above 10$^6$~L$_{\odot}$. Thus,  those GCs that have well-studied multiple populations near SMC
metallicity (e.g.~47~Tuc or M71 with [Fe/H]~$\sim$~$-$0.7) may be explained with our scenario too. The fact that we currently do not see any luminous SG stars in the SMC is not surprising, given the IMF, short lifetime of these stars, and low starformation rate in the SMC. 

\subsection{Proposing another solution for the mass budget: a non-classical IMF for the second generation}\label{sec:2GIMF}

Discussing the mass budget in Sect.~\ref{sec:massb} and after, we assumed that the mass-distribution of the shell-stars is the same as that of the first generation of low-mass stars between 0.1$-$0.8~M$_{\odot}$, and showed that we need a top-heavy IMF for our shell-scenario to work under this assumption. We did this because it helps to compare our scenario to others, such as the fast rotating stars or the massive binary polluters. However, there is another way around the mass budget problem---one that is unique to our scenario. 

Observationally, it is not excluded that all GC stars with M~$<$~0.6~M$_{\odot}$ are first generation stars (the abundances are always determined near the turn-off, i.e.~at 0.8~M$_{\odot}$). Other scenarios usually do not account for this, as this would make their mass budget solution even more speculative. Indeed, if star formation happens out of the interstellar material in the cluster center, it is already hard to justify why the second generation only harbours stars below 0.8~M$_{\odot}$ and nothing above \citep[as done, for example, in][]{deMink:2009}. It would be even more difficult to explain why the IMF would be truncated at both the high and the low ends. Or why, for that matter, the form of the distribution would not follow the classical power-law observed everywhere in the Universe.

In our shell-scenario however, the mode of star formation is so unusual that the IMF must be quite irregular. Apart from massive stars being justifiably excluded on quite robust grounds, it is not clear whether or not the minimum mass could be even larger than the Bonnor-Ebert mass (0.2~M$_{\odot}$, as quoted in Eq.~\ref{eq:BEmass}). After all, the proto-stars may be still accreting some more mass from the shell. 

So for us, it is fathomable to account for a first generation well above 0.2~M$_{\odot}$, the lower limit depending on the accretion rates of the proto-stars. As an example, if the range to account for was only between 0.6$-$0.8~M$_{\odot}$, then the stars represent 7\% of the total cluster mass (following the classical IMF in Eq.~(\ref{eq:imf})). SGs represent 10\%, but their material is inserted into the second generation of shell-stars with an ill-established efficiency $\xi$. This efficiency was taken to be 100\% in the weak case in Sect.~\ref{sec:Salpeter} and 12\% in the strong case in Sect.~\ref{sec:topheavy}, but we expect its realistic value to be somewhere between. Supposing for example that $\xi$~$=$~70\%, the mass budget is solved with having a first generation as numerous (50:50) as the second generation (10\%~$\times$~$\xi$~$=$~7\%). (With $\xi$~$=$100\%, we get a 60:40 ratio of first vs. second generation, cf.~Sect.~\ref{sec:verymassive}.)

Furthermore, we have no reason to suppose that the form of the mass distribution of the second generation is identical to that of the first generation. It certainly needs further investigations (possibly, 3-dimensional simulations of star formation in a spherical shell) to know more about its supposed mathematical form, but in the most optimistic case where all second generation stars form with 0.6~M$_{\odot}$, the efficiency of inserting SG mass into shell-stars can be as low as $\xi$~$=$~40\% to solve the mass budget with a second generation as numerous as the first (50:50). We can also explain very massive clusters where the ratio is more extreme, cf.~Sect.~\ref{sec:verymassive}, if we suppose larger $\xi$ values.

We recall from Sect.~\ref{sec:topheavy} that $\xi$ depends on three astrophysical effects: the mass loss rate of very massive SGs, the amount of material captured in the shell, and the ratio of TWUIN stars vs. SG stars. All three  are poorly constrained at this point, so it is quite conceivable that their interplay adds up to $\xi$~$\gtrsim$~40\%. 

Note that in these considerations, the mass distribution of the \textit{first} generation of stars (both massive and low-mass) follows the classical (not top-heavy) IMF given in Eq.~(\ref{eq:imf}). Solving the mass budget problem this way---having a justifiably irregular IMF for the \textit{second} generation---is a unique feature of our starforming supergiant shell scenario. 


\subsection{Uncertainties of the star-forming shell scenario}

From the point of view of observations, there is some uncertainty as to whether these massive and cool supergiants with low-metallicity actually exist in nature. This will be addressed in the near future by infrared observations of a larger sample of low-metallicity galaxies by the James Webb Space Telescope. From a theoretical point of view, the physics of these stars with inflated envelopes is quite uncertain, and it is undergoing intensive investigation at the moment \citep{Sanyal:2015,Sanyal:2017}. 
Additionally, as mentioned in Sect.~\ref{sec:evolution}, the mass-loss prescription we use involves an extrapolation beyond the mass range where it has been measured.

The process of star formation in a shell is considered rather delicate. It requires several astrophysical effects to combine: that sufficiently dense and long-lived photoionization-confined shells form around very massive SG stars isotropically, so that gravitational instability could occur and lead to the formation of a second generation of stars. As for the mass budget, either the IMF of the cluster should have an index between $-$1.71 and $-$2.07 (as explained in Sect.~\ref{sec:massb}---also note that the upper limit for the first generation, 0.8~M$_{\odot}$ is rather arbitrary), or the second generation should have a non-classical IMF, truncated at both the high and the low end. Additionally, massive stars in this cluster should have a broad rotational velocity distribution, because the TWUIN stars that produce the ionizing radiation are fast rotators. 
Under these conditions, the star-forming shell scenario could potentially produce a second population of stars with the observed abundance variations, and with a similar total mass to that of the first generation of low-mass stars.

\subsection{Supergiants may end up as massive black holes in globular clusters}

With the direct detection of merging black holes via their gravitational wave radiation \citep{Abbott:2016b,Abbott:2016a,Bagoly:2016,Abbott:2017,Szecsi:2017}, many authors suggested globular clusters as the host of these black holes \citep{Rodriguez:2015,Antonini:2016,Belczynski:2016,Askar:2017}. In this section, we discuss the final fate and remnants of our supergiant models.

The cores of very massive stellar models at low-Z undergo pair-instability \citep{Burbidge:1957,Langer:1991,Heger:2003,Langer:2007,Yoon:2012,Kozyreva:2014}. This instability makes the core collapse during oxygen burning, that is, before an iron-core could form. Above a helium-core mass of $\sim$133~M$_{\odot}$, the collapse directly leads to black hole formation. Below this mass, however, it leads to a pair-instability supernova \citep{Heger:2002}. 

From the three supergiant models presented in the context of the star-forming supergiant shell scenario, the most massive two models (with M$_{\rm ini}$=575~M$_{\odot}$ and M$_{\rm ini}$=257~M$_{\odot}$) are predicted to form black holes \textsl{without} a supernova explosion \citep{Szecsi:2016}. 
The masses of these black holes are expected to be above 140~M$_{\odot}$, depending on the strength of the mass-loss (discussed in Sect.~\ref{sec:evolution}). They will thus contribute to the black hole population of their globular clusters. 

The model with M$_{\rm ini}$=150~M$_{\odot}$ on the other hand, which has a final mass of 118~M$_{\odot}$, is predicted to explode as a pair-instability supernova \citep{Szecsi:2016}. The explosion of the SG star may disrupt the shell, but leave the majority of the proto-stars intact. The supenova ejecta is probably too energetic to stay in the cluster's potential well \citep{Lee:2009}, so it may not pollute the second generation of stars \citep[cf.~however][]{Wunsch:2016}. 

\ 

\section{Conclusions}\label{sec:conclusionshell}

We propose star-forming shells around cool supergiants as a possible site to form the second generation of low-mass stars in Galactic globular clusters at low-metallicity. Photoionizaton-confined shells around core-hydrogen-burning cool supergiant stars may have been common in the young GCs. We simulate such a shell and find that it is dense enough to become gravitationally unstable. The new generation of low mass stars that would be formed in the shells should have an initial composition reflecting that of the supergiant's stellar wind, i.e.\ polluted by hot-hydrogen-burning products. 

We summarize the most important ingredients of our star-forming shell scenario below.

\begin{enumerate}

\item \textbf{Low-metallicity supergiant models.} We present state-of-the-art stellar models of low-metallicity supergiants. At this low-metallicity (comparable to the metallicity of globular clusters), our models spend several hundreds of thousands of years on the supergiant branch already during their core-hydrogen-burning phase. They also stay on the supergiant branch during their remaining evolution. 

\item \textbf{Slow, but strong stellar wind.} The supergiant models lose a significant amount of their material in their winds. Since the winds are slow, the material likely stays inside the young globular cluster.

\item \textbf{Hot-hydrogen burning.} In our models of very massive supergiants, the two nuclear burning cylces (Ne-Na chain and Mg-Al chain) that are responsible for the anticorrelation (of O~vs.~Na and Mg~vs.~Al, respectively) are effective. 

\item \textbf{Convective envelope even during hydrogen-burning.} Although the burning processes take place in the core during the core-hydrogen-burning phase, the ashes are mixed between the core and the surface due to the large convective envelope of the supergiant. The composition of the stellar wind is, therefore, enhanced in Na and Al, while depleted of O and Mg. 

\item \textbf{Presence of ionizing sources (TWUIN stars).} We point out that in a population of low-Z massive stars with a broad rotational velocity distribution, the fastest rotating stars will evolve quasi-homogeneously. This chemically homogeneous evolution is responsible for the creation of hot, luminous objects with intense ionizing ratiation, the so-called Transparent Wind UV-Intense stars. We suppose that the radiation field of TWUIN stars is approximately isotropic in the young globular cluster. 

\item \textbf{Photoionization-confined shells.} Where the neutral, cool stellar wind of the supergiants meet the ionized, hot region of the cluster environment, photoionization-confined shells may form. We simulate such a photoionization-confined shell around one of our supergiant models. The shell has a density of 2$\times$10$^{-16}$~g~cm$^{-3}$ and temperature of $\sim$50~K.

\end{enumerate}

%
%

We analyse the stability of the photoionization-confined shell in our simulation, and find that it is gravitationally unstable on a timescale much shorter than the lifetime of the supergiant. The Bonnor-Ebert mass of the overdense regions is low enough to allow star formation. The mass distribution of the new stars is unknown, but we certainly expect the majority of them to be above 0.2~M$_{\odot}$ and below 1~M$_{\odot}$. It is unlikely that massive stars would form because of the geometry of this particular star-forming region.

We show that the composition of a star formed in the photoionization-confined shell is comparable to the observed composition of old, low-mass stars in the most extremely polluted population in globular clusters. We match the abundances of O, Na, Al and Mg, as well as the isotopes of $^{24}$Mg, $^{25}$Mg and $^{26}$Mg. We emphasize that the very high masses of our supergiant models naturally explain the Mg isotope observations, with which some of the alternative scenarios (fast rotating star scenario and the massive binary scenario) clearly struggle. Our scenario also
only works in metal-poor environments however, and cannot apply to the most metal-rich clusters.

Our simulated shell-stars have a high surface helium mass fraction of Y$_\mathrm{sh}$=0.48. We find that low-metallicity supergiants behave the same way as other massive polluters when it comes to helium: they can also only explain the spread in Na\&O and Mg\&Al together with a high helium abundance. But this issue is generic, as both the Ne-Na chain and the Mg-Al chain are side reactions of hot hydrogen-burning \citep{Bastian:2015,Lochhaas:2017}. 

To fulfill the mass-budget constraint, we offer two possibilities. One possibility is that we apply a top-heavy initial mass function with an index being somewhere between $-$1.71 and $-$2.07. These values are less restrictive than those required for some of the other scenarios; e.g. the supermassive stars with 10$^4$~M$_{\odot}$ of \citet{Denissenkov:2014} or the fast rotating stars of \citet{Decressin:2007}. Another possibility is to use a non-classical IMF for the second generation of stars in the shell. We argued that both massive stars and very low-mass stars are justifiably excluded from this second generation, making possible for us to solve the mass budget by accounting only for a fraction of the first generation low-mass stars. 

We emphasize that even if the shells are destroyed e.g. by collision, the corresponding gas may sink into the cluster core and lead to star formation there. Thus, supergiant shells should be considered possible contributors to the chemical evolution of globular clusters.

If the conditions do not facilitate the formation of a photoionization-confined shell (e.g.~because the ionizing radiation field is too weak), the supergiant stellar models presented here should still be considered. Their winds are slow, strong and enhanced by ashes of hot-hydrogen burning. Therefore, our low-Z supergiant models should be taken into account when one is assessing all the possible sources of pollution in young globular clusters. 


Although there are some uncertainties necessarily associated with our proposed scenario of star-forming shells around cool supergiant stars, it shows strong potential for explaining at least some of the second generation of stars with anomalous abundances in GCs -- especially the more extreme cases. Our calculations presented here show that the cool supergiant scenario, both with or without a photoionization-confined shell, deserves serious consideration alongside other, more established scenarios, and should be investigated in more detail in the future.


\begin{acknowledgements}
We thank S.E. de~Mink for her useful comments on the issues of helium spread and the initial composition of the clusters. We also thank R. Wünsch for the careful reading and commenting of our draft, and for his contribution to the discussion of collision times. For the original version of our Fig.~\ref{fig:shell}, we acknowledge its creator, S. Mohamed. 
D.Sz.\ was supported by the Czech Grant nr.\ 13-10589S GA \v{C}R.
JM acknowledges funding from a Royal Society--Science Foundation Ireland University Research Fellowship. This research was partially supported by STFC.
\end{acknowledgements}


\bibliographystyle{aa} 
\bibliography{References} 

\end{document}